\documentclass[12pt, draftclsnofoot, onecolumn]{IEEEtran}
\usepackage{subfigure,cite,graphicx,amsmath,amssymb,mathrsfs,epsfig,dblfloatfix,epstopdf}
\usepackage{amsmath,amssymb,amsthm,mathrsfs,graphicx}
\usepackage{psfrag}
\usepackage{multirow}
\usepackage{array,booktabs}
\usepackage{xcolor}

\newtheorem{theorem}{Theorem}
\newtheorem{example}{Example}
\newtheorem{remark}{Remark}
\newtheorem{lemma}{Lemma}

\newtheorem{corollary}{Corollary}

\begin{document}
\title{Degrees of Freedom of Full-Duplex Cellular Networks with Reconfigurable Antennas at Base Station}

\author{Minho~Yang,~\IEEEmembership{Student~Member,~IEEE,}
				Sang-Woon~Jeon,~\IEEEmembership{Member,~IEEE,}
				and~Dong~Ku~Kim,~\IEEEmembership{Senior~Member,~IEEE}
\thanks{M. Yang and D. K. Kim are with the School of Electrical and Electronic Engineering,
Yonsei University, Seoul, South Korea (e-mail:
\{navigations, dkkim\}@yonsei.ac.kr).}
\thanks{S.-W. Jeon is with the Department of Information and Communication Engineering, Andong National University, Andong, South Korea (e-mail:
swjeon@anu.ac.kr).}
}
\maketitle

\IEEEpeerreviewmaketitle
\begin{abstract}
Full-duplex (FD) cellular networks are considered in which a FD base station (BS) simultaneously supports a set of half-duplex (HD) downlink (DL) users and a set of HD uplink (UL) users.
The transmitter and the receiver of the BS are equipped with reconfigurable antennas, each of which can choose its transmit or receive mode from several preset modes. 
Under the no self-interference assumption arisen from FD operation at the BS, the sum degrees of freedom (DoF) of FD cellular networks is investigated for both no channel state information at the transmit side (CSIT) and partial CSIT.
In particular, the sum DoF is completely characterized for no CSIT model and an achievable sum DoF is established for the partial CSIT model, which improves the sum DoF of the conventional HD cellular networks.
For both no CSIT and partial CSIT models, the results show that the FD BS with reconfigurable antennas can double the sum DoF even in the presence of user-to-user interference as both the numbers of DL and UL users and preset modes increase.
It is further demonstrated that such DoF improvement indeed yields the sum rate improvement at the finite and operational signal-to-noise ratio regime.
\end{abstract}

\begin{IEEEkeywords}
Blind interference alignment, degrees of freedom (DoF), full-duplex (FD), interference management, reconfigurable antennas.
\end{IEEEkeywords}

\section{Introduction}
To meet soaring wireless demand with limited spectrum, there has been considerable researches for boosting utilization of wireless resources.
Recently, \emph{full-duplex (FD) radios} have emerged as a potential way of improving spectral efficiency by enabling simultaneous transmission and reception at the same time with the same wireless spectrum. 
Because of such simultaneous transmission and reception, FD has a potential to double the spectral efficiency compared to the conventional half-duplex (HD) mode such as frequency division duplex (FDD) and time division duplex (TDD).
Nonetheless, FD involves the practical issue of suppressing high-powered self-interference arisen from simultaneous transmission and reception \cite{Hong:14, Duarte:12, Bharadia:13, Chung:15}.
In recent researches, there has been remarkable progress on analog and digital domain self-interference cancellation (SIC) techniques, showing that the point-to-point bidirectional FD system can achieve nearly twice higher throughput than the corresponding HD system, which demonstrates the possibility of implementing FD radios in practice \cite{Duarte:12, Bharadia:13, Chung:15}.

\begin{figure}[!t] 
\includegraphics[width=2.5in]{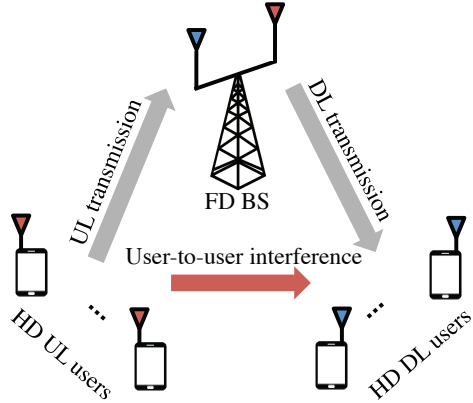}
\centering 
\caption{User-to-user interference for FD cellular networks.} 
\label{FD_cellular}
\end{figure}

Unlike the point-to-point bidirectional FD system, we cannot simply argue that the network throughput can be doubled for cellular systems even under the ideal assumption that self-interference is perfectly suppressed.
In particular, consider the cellular system in Figure \ref{FD_cellular} in which a FD base station (BS) simultaneously supports a set of HD downlink (DL) users and a set of HD uplink (UL) users, one of the feasible scenarios of FD radios considering compatibility with legacy HD users in the current communication systems. For such case, a new source of interference from UL users to DL users appears, which does not exist in HD cellular systems where DL and UL traffic is orthogonalized by frequency or time domain.
The impact of such user-to-user interference in FD cellular systems has been widely discussed in several researches \cite{Sabharwal:14, Kim:15, Goyal:15,Jeon:14,KimJeon:14}.
They showed that if interference from UL users  to DL users is not properly mitigated, the network throughput may be degraded even though self-interference is perfectly suppressed.
Therefore, efficient interference management from UL users to DL users is a key challenge to boosting the network throughput of cellular systems by adapting FD operation at BSs \cite{Sabharwal:14, Kim:15, Goyal:15,Jeon:14,KimJeon:14}.

In order to understand fundamental limits of FD radios in cellular networks, there have been several recent researches on characterizing the {\it degrees of freedom (DoF) of FD cellular networks} \cite{Sahai13,chae14, Jeon:15, Bai:15}.
In particular, a single-cell FD cellular network has been studied in \cite{Jeon:15, Bai:15}, in which a FD BS with perfect self-interference suppression supports both HD DL and UL users as seen in Fig. \ref{FD_cellular}.
%
%
In \cite{Jeon:15}, the authors characterized the sum DoF of the single-cell FD cellular network assuming that global channel state information (CSI) is available at the BS, i.e., full CSI at the transmit side (CSIT).
They showed that FD operation at the BS can double the sum DoF compared to HD operation when both the numbers of DL and UL users become large even in the presence of user-to-user interference, concurrently reported in \cite{Bai:15}.
However, asymptotic interference alignment (IA) techniques proposed in \cite{Jeon:15,Bai:15} require perfect CSIT and an arbitrarily large number of time extension to achieve the optimal sum DoF,
which is quite challenging in practice due to feedback delay, system overhead and complexity, and etc \cite{Cadambe:08,Suh:11,Viveck1:09,Viveck2:09,Tiangao:10,Tiangao:12,Jeon4:12}. 

To resolve such practical restrictions for interference management, the concept of \emph{blind IA} has been recently proposed, which aligns multiple interfering signals into the same signal space at each receiver without any CSIT.
In particular, various blind IA techniques have been proposed for both heterogeneous block fading models where certain users experience smaller coherence time/bandwidth than others \cite{Jafar:12} and homogeneous block fading models where all users experience independent block fading with the same coherence time, but different offsets \cite{Zhou:12, Zhou:12_dio, Zhou:12_Imp}.
In \cite{Wang:10}, Wang, Gou, and Jafar have first observed that \emph{reconfigurable antennas} can artificially create channel correlation across time in a certain structure letting blind IA be possible for  multiple-input and multiple-output (MIMO) broadcast channels \cite{Gou:11, Wang:10}.
Reconfigurable antennas are capable of dynamically adjusting its radiation patterns in a controlled and reversible manner through various technologies such as solid state switches or microelectromechanical switches (MEMS) without additional RF-chains, which take a dominant factor for hardware complexity \cite{Christ:12, Sohn:16}. That is, reconfigurable antennas can choose its transmit or receive mode among several preset modes at each time instant, see also \cite[Section I]{Gou:11} for the concept of reconfigurable antennas.
Subsequently, blind IA using reconfigurable antennas has been extended to general MIMO broadcast channels characterizing linear sum DoF, i.e., the maximum sum DoF achievable by linear coding schemes \cite{Yang:14} and also applied to a class of single-input and single-output (SISO) and multiple-input and single-output (MISO) interference channels consisting of receivers equipped with reconfigurable antennas \cite{Wang:14, Lu:14}.
From the recent results in \cite{Gou:11, Wang:10,Yang:14,Wang:14, Lu:14} together with the advantage of reconfigurable antennas on hardware complexity \cite{Christ:12, Sohn:16}, blind IA using reconfigurable antennas has been considered as a promising solution for boosting the DoF of practical wireless systems with no CSIT.

Motivated by such advantages of FD radios and reconfigurable antennas, 
we consider FD cellular networks in which a FD BS equipped with reconfigurable transmit and receive antennas supports HD DL and UL users simultaneously in the same frequency spectrum.
For comprehensive understanding on the impact of FD radios and CSI conditions in the context of IA or blind IA using reconfigurable antennas, we consider two different CSI models: For no CSIT case, both the BS and each UL user do not know their CSIT; For the partial CSIT case, the BS only knows its CSIT. For both models, we assume that CSI at the receive side (CSIR) is available.
Similar to the previous full CSIT models in \cite{Sahai13,chae14, Jeon:15, Bai:15}, the primary aim is to characterize whether the sum DoF can be doubled or not with partial or no CSIT by FD operation at the BS equipped with reconfigurable antennas. The main contributions of this paper are as follows:
\begin{itemize}
\item For no CSIT model, we completely characterize the sum DoF of FD cellular networks. We propose a novel blind IA technique, which perfectly aligns user-to-user interference at each DL user while preserving intended signal space at the BS, and establish the converse showing the optimality of the proposed scheme in terms of the sum DoF. The result shows that the sum DoF is asymptotically doubled if both the numbers of UL users and preset modes at the receiver of the BS increase, which is the first result demonstrating the benefit of FD radios on cellular networks under no CSIT.
\item For the partial CSIT model, we establish an achievable lower bound on the sum DoF of FD cellular networks,
which characterizes the sum DoF for a broad class of network topologies.
We propose a novel blind IA technique combined with zero-forcing beamforming based on partial CSIT, which partially aligns user-to-user interference at each DL user while preserving intended signal space at the BS.
 The result shows that the sum DoF is doubled if there exist two DL and two UL users and two preset modes at the transmitter and the receiver of the BS. For the single-antenna case, our result for the partial CSIT model extends the previous achievability result in \cite{Bai:15} to a general antenna configuration assuming different numbers of preset modes at the transmitter and receiver of the BS.
\item We further demonstrate that such DoF improvement indeed yields the sum rate improvement at the finite and operational signal-to-noise ratio (SNR) regime, which presents the benefit of blind IA using reconfigurable antennas compared with the previous works \cite{Sahai13,chae14, Jeon:15, Bai:15}.
\end{itemize}

The rest of this paper is organized as follows. In Section \ref{sec:system}, we introduce the network model and DoF metric considered throughout the paper. 
In Section \ref{sec:main}, we state the main results of this paper, the sum DoF of FD cellular networks, and remark several observations possibly deduced from the main results. 
We present achievability and converse proofs of the main results in Section \ref{sec:achievability} and Section \ref{sec:converse} respectively.
We finally conclude in Section \ref{sec:conclusion}.

\section{Problem Formulation}\label{sec:system}

In this section, we introduce FD cellular networks consisting of a FD BS and HD DL and HD UL users and then formally define the sum DoF metric, which will be analyzed throughout the paper.

\subsection{Notation}
For integer numbers $a$ and $b$, $a\setminus b$ and $a|b$ denote the quotient and the remainder respectively when dividing $a$ by $b$.
For integer numbers $a$ and $b$, $[a:b] = \{a, a+1, \cdots, b\}$ when $a \leq b$ and $[a:b] = \emptyset$ when $a>b$.
%
%
%
%
For matrices $\mathbf{A}$ and $\mathbf{B}$, $\mathbf{A}\otimes \mathbf{B}$ is the Kronecker product of $\mathbf{A}$ and $\mathbf{B}$.
For a matrix $\mathbf{A}$, denote the Frobenius norm, transpose, and conjugate transpose of $\mathbf{A}$ by $\|\mathbf{A}\|$, $\mathbf{A}^T$, and $\mathbf{A}^{H}$, respectively.
For a set of matrices $\{\mathbf{A}_{i}\}_{i\in[1:n]}$, $\operatorname{diag}(\mathbf{A}_{1},\cdots,\mathbf{A}_{n})$ denotes the block-diagonal matrix consisting of $\mathbf{A}_{i}$ as the $i$th diagonal block.
For natural numbers $a$ and $b$, $\mathbf{I}_{a}$, $\mathbf{1}_{a\times b}$, and $\mathbf{0}_{a\times b}$ denote the $a \times a$ identity matrix, the $a \times b$ all-one matrix, and the $a \times b$ all-zero matrix respectively.
Let $\mathbf{e}_{a}(b)$ be the $b$th column vector of $\mathbf{I}_{a}$ where $b\in[1:a]$.

\begin{figure}[!t] 
\includegraphics[width=2.5in]{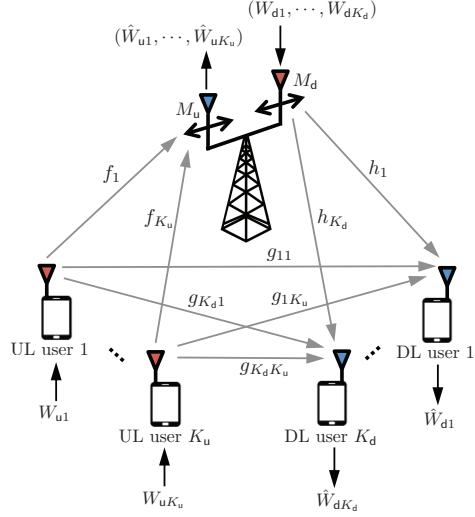}
\centering 
\caption{Full-duplex cellular networks.} 
\label{system}
\end{figure}

\subsection{Full-Duplex Cellular Networks} \label{sec:system1}

We consider a FD cellular network in which a FD BS simultaneously supports $K_{\sf d}$ HD DL users and $K_{\sf u}$ HD UL users.
%
%
Both the transmitter and receiver of the BS are equipped with reconfigurable antennas.
In particular, the transmitter of the BS is equipped with a reconfigurable antenna capable of switching among $M_{\sf d}$ preset modes at each time
and the receiver of the BS is equipped with a reconfigurable antenna capable of switching among $M_{\sf u}$ preset modes at each time.
Notice that $M_{\sf d}=1$ (or $M_{\sf u}=1$) corresponds to the case where the transmitter (or the receiver) of the BS is equipped with a conventional antenna.
Each DL and UL user is equipped with a conventional antenna.
In this paper, we assume that self-interference within the BS due to FD operation is perfectly suppressed. We will discuss about the impact of imperfect self-interference suppression in Section \ref{sec:sum_rates}.

We assume block fading in this paper, i.e., each channel coefficient remains the same in a consecutive time slots of coherence time and is drawn independently in the next consecutive time slots of coherence time.
The length of the coherence time is assumed to be sufficiently large.
Let $h_{i}(k) \in \mathbb{C}$ be the channel from the transmitter of the BS to the $i$th DL user when the BS selects its transmit mode as the $k$th preset mode, where $i\in[1:K_{\sf d}]$ and $k \in [1: M_{\sf d}]$.
Similarly, let $f_{j}(l)\in\mathbb{C}$ be the channel from the $j$th UL user to the receiver of the BS when the BS selects its receive mode as the $l$th preset mode, where $j\in[1:K_{\sf u}]$ and $l \in [1: M_{\sf u}]$.
Let $g_{ij} \in \mathbb{C}$ be the channel from the $j$th UL user to the $i$th DL user.
All channel coefficients are assumed to be independent and identically distributed (i.i.d.) drawn from a continuous distribution.

Denote the transmit mode and the receive mode of the BS at time $t$ by $\alpha(t) \in [1:M_{\sf d}]$ and $\beta(t)\in [1:M_{\sf u}]$, respectively. 
Then the received signal of the $i$th DL user at time $t$ is given by 
\begin{align}
y_{{\sf d}i}(t) & = h_{i}(\alpha(t)) x_{\sf d}(t) + \sum_{j=1}^{K_{\sf u}} g_{ij} x_{{\sf u}j}(t) + z_{{\sf d}i} (t) \label{eq:dl}
\end{align}
for $i\in[1:K_{\sf d}]$
and the received signal of the BS at time $t$ is given by
\begin{align}
y_{\sf u}(t) & =  \sum_{j=1}^{K_{\sf u}} f_{j}(\beta(t)) x_{{\sf u}j}(t) + z_{\sf u} (t) \label{eq:ul}
\end{align}
where 
$x_{\sf d}(t)$ is the transmit signal of the BS at time $t$, 
$x_{{\sf u}j}(t)$ is the transmit signal of the $j$th UL user at time $t$, $z_{{\sf d}i}(t)$ is the additive noise of the $i$th DL user at time $t$, and $z_{\sf u}(t)$ is the additive noise of the BS at time $t$. 
The additive noises are assumed to be i.i.d. drawn from $\mathcal{CN}(0,1)$ and independent over time.
The BS and each UL user should satisfy the average power constraint $P$, i.e., 
$\mathbb{E} \left[ \|x_{\sf d}(t)\|^2 \right ]   \leq P$ and  $\mathbb{E} \left[ \|x_{{\sf u}j}(t)\|^2 \right ]  \leq P$ for all $ j\in [1:K_{\sf u}]$.

For notational convenience, from \eqref{eq:dl} and \eqref{eq:ul}, we define the length-$n$ time-extended input--output relation as
\begin{align} \label{eq:in_out_extended}
				\mathbf{y}_{{\sf d}i} & = \mathbf{H}_{i}(\bar{\alpha}) \mathbf{x}_{\sf d} + \sum_{j=1}^{K_{\sf u}} g_{ij} \mathbf{x}_{{\sf u}j}  + \mathbf{z}_{{\sf d}i}, \notag \\
\mathbf{y}_{\sf u} & = \sum_{j=1}^{K_{\sf u}} \mathbf{F}_{j}(\bar{\beta}) \mathbf{x}_{{\sf u}j} + \mathbf{z}_{\sf u}
\end{align}
where
\begin{align*}
 \bar{\alpha} & =[\alpha(1),\cdots,\alpha(n)]^T,\ \bar{\beta}=[\beta(1),\cdots,\beta(n)]^T,\\
 \mathbf{H}_i(\bar{\alpha}) & = \operatorname{diag}\left( h_i(\alpha(1)), \cdots, h_i(\alpha(n)) \right), \\ 
 \mathbf{F}_j(\bar{\beta})  & = \operatorname{diag}\left( f_j(\beta(1)), \cdots, f_j(\beta(n)) \right),  \\ 
 \mathbf{y}_{{\sf d}i} & = \left[y_{{\sf d}i}(1), \cdots, y_{{\sf d}i}(n) \right]^T, \ \mathbf{y}_{\sf u} = \left[y_{\sf u}(1), \cdots, y_{\sf u}(n) \right]^T,\\
 \mathbf{x}_{\sf d} & = \left[x_{\sf d}(1),\cdots,x_{\sf d}(n)\right]^T, \ \mathbf{x}_{{\sf u}i} = \left[x_{{\sf u}i}(1),\cdots,x_{{\sf u}i}(n)\right]^T, \\
 \mathbf{z}_{{\sf d}i} & = \left[z_{{\sf d}i}(1), \cdots, z_{{\sf d}i}(n)\right]^T, \ \mathbf{z}_{{\sf u}}	= \left[z_{{\sf u}}(1), \cdots, z_{{\sf u}}(n)\right]^T.
\end{align*}

For comprehensive understanding on the DoF improvement achievable by reconfigurable antennas at the FD BS, we consider the following two different scenarios for CSI assumption:
\begin{itemize}
\item {\bf No CSIT model} (CSIT is not available):\\
The BS knows its receive side CSI, $\{f_{j}(k)\}_{j\in[1:K_{\sf u}],k\in[1:M_{\sf u}]}$; The $i$th DL user knows its receive side CSI, $\{h_{i}(k)\}_{k\in[1:M_{\sf d}]}$;
The $j$th UL user does not know any CSI.
\item {\bf Partial CSIT model} (CSIT is only available at the BS):
\\The BS knows both its transmit and receive side CSI, i.e., $\{h_{i}(k)\}_{i\in[1:K_{\sf d}],k\in[1:M_{\sf d}]}$ and $\{f_{j}(k)\}_{j\in[1:K_{\sf u}],k\in[1:M_{\sf u}]}$; The $i$th DL user knows its receive side CSI, $\{h_{i}(k)\}_{k\in[1:M_{\sf d}]}$; The $j$th UL user does not know any CSI.
\end{itemize}

\begin{remark}
For the considered network, CSIR might not immediately lead to CSIT even if channel reciprocity holds because a FD BS supports HD DL users and HD UL users. 
That is, a set of DL users and a set of UL users are fixed and separate.
Furthermore, the validity of such channel reciprocity will depend on the relative difference between channel coherence time and time difference between UL and DL frames allocated to an user. 
If the time difference between UL and DL frames allocated to an user is longer than the coherence time, then additional channel feedback from the receive side to the transmit side is required to attain CSIT \cite{Kim:14}.
Moreover, the RF front-ends of transmit and receive antennas are different and have their own delays and gains,
which necessarily cause reciprocity error and impose reciprocity calibration \cite{Shi:11}.
For the above reasons, we consider both no CSIT and partial CSIT models in this paper. \hfill$\lozenge$
\end{remark}

\begin{remark}
Notice that, for both no CSIT and partial CSIT models in this paper, each DL user does not require CSI from its UL users. Therefore, 
CSIR is available by using the conventional UL channel training (for CSI from UL users to the BS) and DL channel training (for CSI from the BS to DL users) without additional channel training from UL to DL users. 
 \hfill$\lozenge$ 
\end{remark}

\subsection{Degrees of Freedom} \label{sec:DoF}
For the network model stated in Section \ref{sec:system1}, we define a set of length-$n$ block codes and its achievable DoF.
Let $W_{{\sf d}i}\in [1: 2^{nR_{{\sf d}i}} ]$ and $W_{{\sf u}j}\in [1: 2^{nR_{{\sf u}j}} ]$ be the $i$th DL message and the $j$th UL message respectively, where $i \in [1:K_{\sf d}]$ and $j \in [1:K_{\sf u}]$.
For no CSIT model, a $(2^{nR_{{\sf d}1}},\cdots,2^{nR_{{\sf d}K_{\sf d}}},2^{nR_{{\sf u}1}},\cdots,2^{nR_{{\sf u}K_{\sf u}}};n)$ code consists of the following set of encoding and decoding functions:
\begin{itemize}
\item Encoding: For $t \in [1:n]$, the encoding function of the BS at time $t$ is given by
\begin{align}
\left ( x_{\sf d}(t), \alpha(t) \right ) = \phi_t\left(W_{{\sf d}1},\cdots,W_{{\sf d}K_{\sf d}}, y_{\sf u}(1),\cdots, y_{\sf u}(t-1), \{f_{j}(k)\}_{j\in[1:K_{\sf u}],k\in[1:M_{\sf u}]} \right ). \notag
\end{align}

For $t \in [1:n]$, the encoding function of the $j$th UL user ($j\in[1:K_{\sf u}]$) at time $t$ is 
\begin{align}
x_{{\sf u}j}(t) = \varphi_{jt}\left(W_{{\sf u}j} \right ). \notag
\end{align}

\item Decoding:
Upon receiving $\mathbf{y}_{\sf u}$ (i.e., $y_{\sf u}(1)$ to $y_{\sf u}(n)$), the decoding function of the BS is 
\begin{align}
\hat{W}_{{\sf u}j} = \chi_j \left(\mathbf{y}_{\sf u}, W_{{\sf d}1}, \cdots, W_{{\sf d}K_{\sf d}}, \{f_{j}(k)\}_{j\in[1:K_{\sf u}],k\in[1:M_{\sf u}]} \right ) \mbox{ for } j \in [1:K_{\sf u}]. \notag
\end{align}

Upon receiving $\mathbf{y}_{{\sf d}i}$, 
the decoding function of the $i$th DL user ($i\in[1:K_{\sf d}]$) is given by

\begin{align}
\hat{W}_{{\sf d}i} = \psi_i \left(\mathbf{y}_{{\sf d}i}, \{h_{i}(k)\}_{k\in[1:M_{\sf d}]} \right ). \notag
\end{align}
\end{itemize}

If there exists a sequence of $(2^{nR_{{\sf d}1}},\cdots,2^{nR_{{\sf d}K_{\sf d}}},2^{nR_{{\sf u}1}},\cdots,2^{nR_{{\sf u}K_{\sf u}}};n)$ codes such that $\mbox{Pr}(\hat{W}_{{\sf d}i} \neq W_{{\sf d}i}) \rightarrow 0$ and $\mbox{Pr}(\hat{W}_{{\sf u}j} \neq W_{{\sf u}j}) \rightarrow 0$ as $n$ increases for all $i \in [1:K_{\sf d}]$ and $j \in [1:K_{\sf u}]$, a rate tuple $(R_{{\sf d}1}, \cdots, R_{{\sf d}K_{\sf d}}, R_{{\sf u}1},\cdots, R_{{\sf u}K_{\sf u}})$ is said to be achievable.
Then the achievable DoF tuple is given by
\begin{align}
(d_{{\sf d}1},\cdots, d_{{\sf d}K_{\sf d}},d_{{\sf u}1},\cdots, d_{{\sf u}K_{\sf u}}) = \underset{P \rightarrow \infty }{\lim} 
\left (\frac{R_{{\sf d}1}}{\log P}, \cdots, \frac{R_{{\sf d}K_{\sf d}}}{\log P}, \frac{R_{{\sf u}1}}{\log P},\cdots, \frac{R_{{\sf u}K_{\sf u}}}{\log P}\right). \notag
\end{align}
Finally, the sum DoF for no CSIT model is defined as
\begin{align}
d_{\Sigma, {\sf no CSIT}} = \underset{(d_{{\sf d}1},\cdots, d_{{\sf d}K_{\sf d}},d_{{\sf u}1},\cdots, d_{{\sf u}K_{\sf u}}) \in \mathcal{D}}{\max}  
\left \{  \sum\limits_{i=1}^{K_{\sf d}} d_{{\sf d}i} + \sum\limits_{j=1}^{K_{\sf u}} d_{{\sf u}j}\right \} \notag
\end{align}
where $\mathcal{D}$ denotes the achievable DoF region.

For the partial CSIT model, the encoding and decoding functions of the BS are replaced as 
\begin{align*}
& \left ( x_{\sf d}(t), \alpha(t) \right ) \notag \\
& \ \ \ \ = \phi_t\big(W_{{\sf d}1},\cdots,W_{{\sf d}K_{\sf d}}, y_{\sf u}(1),\cdots,  y_{\sf u}(t-1),\{h_{i}(k)\}_{i\in[1:K_{\sf d}],k\in[1:M_{\sf d}]}, \{f_{j}(k)\}_{j\in[1:K_{\sf u}],k\in[1:M_{\sf u}]} \big ),\\
& \hat{W}_{{\sf u}j} = \chi_j \big(\mathbf{y}_{\sf u}, W_{{\sf d}1}, \cdots, W_{{\sf d}K_{\sf d}},\{h_{i}(k)\}_{i\in[1:K_{\sf d}],k\in[1:M_{\sf d}]}, \{f_{j}(k)\}_{j\in[1:K_{\sf u}],k\in[1:M_{\sf u}]} \big ),
\end{align*}
respectively.
Then the sum DoF can be defined in the same manner. Let $d_{\Sigma, {\sf pCSIT}}$ denote the sum DoF for the partial CSIT model.

For the rest of this paper, we characterize the sum DoF of the FD cellular network under both no CSIT model and the partial CSIT model.


\section{Main Results}\label{sec:main}
In this section, we state our main results, the sum DoF of the FD cellular network for both no CSIT and partial CSIT models, and provide a numerical example for demonstrating the benefit of FD operation and reconfigurable antennas at the BS.

For no CSIT model, we completely characterize the sum DoF of the FD cellular network in the following theorem.
\begin{theorem} \label{theorem:3}
For the FD cellular network with no CSIT, 
\begin{align} \label{eq:DoF11}
d_{\Sigma, {\sf noCSIT} } & =   \min \left \{ \max(K_{\sf d}, K_{\sf u}), \max \left ( 1 +  \frac{\min(K_{\sf d},1)(L_{\sf u}-1)}{L_{\sf u}} , 1 \right ) \right \}
\end{align}
where $L_{\sf u} = \min(K_{\sf u}, M_{\sf u} )$.
\begin{IEEEproof}
We refer achievability proof to Section \ref{proof:theorem1} and converse proof to Section \ref{sec:converse}.
\end{IEEEproof}
\end{theorem}

\begin{remark}
From Theorem \ref{theorem:3}, $d_{\Sigma, {\sf noCSIT} }$ is independent of the parameters $K_{\sf d}$ and $M_{\sf d}$ if $K_{\sf d} \neq 0$ and $K_{\sf u} \neq 0$.
That is, for no CSIT case, equipping a reconfigurable antenna at the transmitter of the BS cannot increase the sum DoF and similarly a single DL user is enough to achieve the optimal sum DoF.
More importantly, \emph{$d_{\Sigma, {\sf noCSIT} }$  is asymptotically doubled if both $K_{\sf u}$ and $M_{\sf u}$ increase.}
Therefore, for no CSIT case, arbitrarily large numbers of UL users and preset modes at the receiver of the BS are required to double the sum DoF by FD operation at the BS.  \hfill$\lozenge$ 
\end{remark}

For the partial CSIT model, we establish an upper and achievable lower bounds on the sum DoF of the FD cellular network in the following theorem.
\begin{theorem} \label{theorem:1}
For the FD cellular network with partial CSIT, 
\begin{align} \label{eq:DoF0}
d_{\Sigma, {\sf pCSIT}}
  \leq \min \left \{
2, \max \left (K_{\sf d}, K_{\sf u} \right ), \max \left ( 1 + \frac{K_{\sf u} (K_{\sf d}-1)}{K_{\sf d}}, 1 + \frac{K_{\sf d} (K_{\sf u}-1)}{K_{\sf u}} \right )
\right \} 
\end{align}
and 
\begin{align} 
d_{\Sigma, {\sf pCSIT}} \geq  \label{eq:DoF} 
 \min \left \{
2, \max(K_{\sf d}, K_{\sf u}) ,\max \left ( 1 + \frac{L_{\sf u}(L_{\sf d}-1)}{L_{\sf d}}, 1 + \frac{L_{\sf d} (L_{\sf u}-1)}{L_{\sf u}} \right ) 
\right \} 
\end{align}
where $L_{\sf d} = \min(K_{\sf d}, M_{\sf d})$ and $L_{\sf u} = \min(K_{\sf u}, M_{\sf u})$.
\begin{IEEEproof}
We refer to the converse in \cite[Theorem 1]{Jeon:15} for the proof of the upper bound in \eqref{eq:DoF0}.
In particular, \cite{Jeon:15} considers the FD BS equipped with conventional multiple transmit and receive antennas (instead of reconfigurable antennas) and assumes that full CSI is available at the BS and each user.
The upper bound in \eqref{eq:DoF0} is attained from \cite[Theorem 1]{Jeon:15} by assuming a single transmit and receive antenna at the BS. We can easily see that the converse argument in \cite[Theorem 1]{Jeon:15} is applicable to the reconfigurable antenna model in Fig. \ref{system} for the full CSIT case. 
Hence \eqref{eq:DoF0} can be an upper bound on $d_{\Sigma, {\sf pCSIT}}$.
We refer to Section \ref{proof:theorem2} for the proof of the achievable lower bound in \eqref{eq:DoF}.
\end{IEEEproof}
\end{theorem}

\begin{corollary} \label{co:sum_dof_pCSIT}
For the FD cellular network with partial CSIT, 
\begin{align} \label{eq:sum_dof_pCSIT}
d_{\Sigma, {\sf pCSIT}}=\begin{cases}
2 &\mbox{if }K_{\sf d},K_{\sf u},M_{\sf d},M_{\sf u}\geq 2,\\
1 + \frac{K_{\sf u}-1}{K_{\sf u}} &\mbox{if } K_{\sf d}=1, M_{\sf u}\geq K_{\sf u}\geq 1,\\
1 + \frac{K_{\sf d}-1}{K_{\sf d}} &\mbox{if } K_{\sf u}=1, M_{\sf d}\geq K_{\sf d}\geq 1.
\end{cases}
\end{align}
\begin{proof}
By comparing the upper and lower bounds on $d_{\Sigma, {\sf pCSIT}}$ in Theorem \ref{theorem:1}, \eqref{eq:sum_dof_pCSIT} can be straightforwardly obtained.
\end{proof}
\end{corollary}

For the single-antenna case, Theorem \ref{theorem:1} and Corollary \ref{co:sum_dof_pCSIT} extend the previous achievability result for the partial CSIT model in \cite{Bai:15} to a general antenna configuration assuming different numbers of preset modes at the transmitter and receiver of the BS.

\begin{remark}
From Theorem \ref{theorem:1} and Corollary \ref{co:sum_dof_pCSIT}, \emph{$d_{\Sigma, {\sf pCSIT}}$ is asymptotically doubled if both $K_{\sf u}$ and $M_{\sf u}$ increase when $\min(K_{\sf d},M_{\sf d})=1$ or both $K_{\sf d}$ and $M_{\sf d}$ increase when $\min(K_{\sf u},M_{\sf u})=1$.}
Hence, similar to no CSIT case, arbitrarily large numbers of users and preset modes are required to double the sum DoF by FD operation only at the DL or UL side.
On the other hand, \emph{$d_{\Sigma, {\sf pCSIT}}$ is doubled if $M_{\sf d},M_{\sf u},K_{\sf d},K_{\sf u}\geq 2$.} That is, only two DL and UL users and the FD BS equipped with reconfigurable antennas having two preset modes are enough to double the sum DoF if the BS can attain its downlink CSI.
Lastly, unlike no CSIT case in which reconfigurable antennas are only beneficial at the receiver of the BS, reconfigurable antennas are equally beneficial at the transmitter and receiver of the BS for the partial CSIT case.  \hfill$\lozenge$ 
\end{remark}

\begin{figure}[!t]
\includegraphics[width=3.4in]{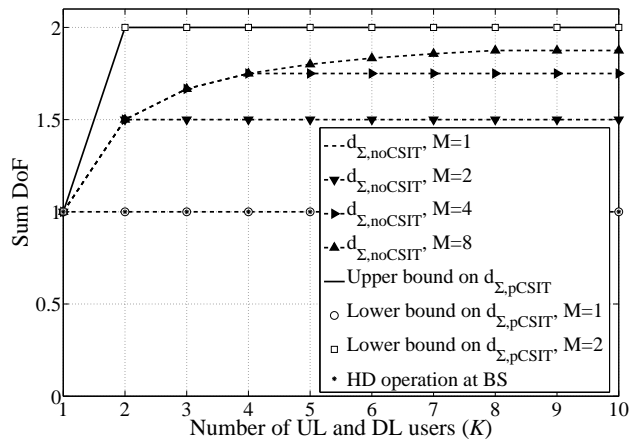}
\centering
\caption{Sum DoFs with respect to $K$ when $K_{\sf d} = K_{\sf u} := K$ and $M_{\sf d} = M_{\sf u} := M$.}
\label{symmetric}
\end{figure}

In summary, from Theorems \ref{theorem:3} and \ref{theorem:1}, the sum DoF is doubled even in the presence of user-to-user interference by FD operation at the BS. Furthermore, reconfigurable antennas can effectively improve the sum DoF under both partial and no CSIT cases. The following example plots the sum DoFs in Theorems \ref{theorem:3} and \ref{theorem:1} for the symmetric case.

\begin{example}
For comparison, consider the symmetric case where $K_{\sf d} = K_{\sf u} := K \geq 1$ and $M_{\sf d} = M_{\sf u} := M$.
Then, from Theorem \ref{theorem:3} and \ref{theorem:1}, 
\begin{align} \label{eq:example1}
d_{\Sigma, {\sf noCSIT}} = 2 - \frac{1}{\min(K,M)}
\end{align}
and
\begin{align}
\min(K,M,2)\leq d_{\Sigma, {\sf pCSIT}} \leq \min(K,2).  \label{eq:example2}
\end{align}
Fig. \ref{symmetric} plots \eqref{eq:example1} and \eqref{eq:example2}  with respect to $K$.
Obviously, if the BS operates as the conventional HD operation, i.e., serving either DL users or UL users, the sum DoF is limited by one.
From \eqref{eq:example1} and the lower bound in \eqref{eq:example2}, the sum DoF is still one if the FD BS is equipped with conventional non-reconfigurable antennas, i.e., $M=1$.
For the partial CSIT case, $K=M=2$ is enough to double the sum DoF. 
On the other hand, arbitrarily large $K$ and $M$ are required to double the sum DoF in the case of no CSIT.
 \hfill$\lozenge$
\end{example}

In Section \ref{sec:sum_rates}, we further demonstrate that the above sum DoF improvement achievable by FD operation and reconfigurable antennas at the BS yields the sum rate at the finite and operational SNR regime, which presents the benefit of blind IA using reconfigurable antennas compared with the previous works \cite{Sahai13,chae14, Jeon:15, Bai:15}.


\section{Achievability}\label{sec:achievability}

In this section, we establish the achievability in Theorems \ref{theorem:3} and \ref{theorem:1} and then present the achievable sum rates of the proposed schemes at the finite SNR regime.

Recall that $L_{\sf d}=\min(K_{\sf d},M_{\sf d})$ and $L_{\sf u}=\min(K_{\sf u},M_{\sf u})$.
When $K_{\sf d} = 0$ or $K_{\sf u} = 0$, the right-hand sides of \eqref{eq:DoF11} and \eqref{eq:DoF} in Theorems \ref{theorem:3} and \ref{theorem:1} are expressed as
\begin{align}
\min \big \{ \max(K_{\sf d}, K_{\sf u}), 1 \big \}. \notag
\end{align}
In this case, the sum DoF is trivially achievable by single-user transmission (supporting a DL user if $K_{\sf d}\neq 0$ and a UL user if $K_{\sf u}\neq 0$).
Thus, we now focus on the achievability proof of Theorems \ref{theorem:3} and \ref{theorem:1} when $K_{\sf d}, K_{\sf u} \geq 1$.

Let us define the $n$-point inverse discrete Fourier transform (IDFT) matrix as $\mathbf{\Omega}_n\in\mathbb{C}^{n\times n}$ 
, given by
\begin{align*}
\mathbf{\Omega}_n = \frac{1}{\sqrt{n}}
\left [
\begin{array}{cccc}
1 			& 1				 			 								& \cdots & 1 										\\
1 			& \omega 				 					& \cdots & \omega^{n-1} 				\\
\vdots 	& \vdots 				 						& \ddots & \vdots 							\\
1				& \omega^{n-1} 	 	& \cdots & \omega^{(n-1)(n-1)} 	\\
\end{array}
\right ]
\end{align*}
where $\omega = e^{j 2 \pi / n}$ \cite{Opp:13}.
In the followng, the IDFT matrix will be used for transmit precoding matrices to exploit the following properties of the IDFT matrix:
1) $\mathbf{\Omega}_n$ is an orthonormal matrix, i.e.,
\begin{align} \label{eq:property_IDFT}
\mathbf{\Omega}_n^{H}\mathbf{\Omega}_n=\mathbf{I}_n;
\end{align}
2) Every submatrix of $\mathbf{\Omega}_n$ is of full-rank \cite{bader2007petascale}. In particular, the above properties will be used to prove Lemma \ref{lemma:dl2}.

\subsection{Achievability for Theorem \ref{theorem:3} when $K_{\sf d}, K_{\sf u} \geq 1$} \label{proof:theorem1}
When $K_{\sf d}, K_{\sf u} \geq 1$, the right-hand side of \eqref{eq:DoF11} is given by
\begin{align}
2 - \frac{1}{L_{\sf u}}. \notag
\end{align}
In the following, we establish the achievability of Theorem \ref{theorem:3}, showing that the sum DoF of $2 - \frac{1}{L_{\sf u}}$ is achievable for no CSIT model.
In particular, the BS sends $L_{\sf u}-1$ information symbols to only the first DL user and $K_{\sf u}$ UL users send a single information symbol each to the BS during $L_{\sf u}$ time slots.

Let $\mathbf{s}_{{\sf d} 1} \in \mathbb{C}^{(L_{\sf u} - 1)\times 1}$ be the information symbol vector for the first DL user satisfying that $\mathbb{E}[\|\mathbf{s}_{{\sf d}1}\|^2] = L_{\sf u}P$ and $s_{{\sf u} j} \in \mathbb{C}$ be the information symbol for the $j$th UL user, $j\in[1:K_{\sf u}]$, satisfying that $\mathbb{E}[|s_{{\sf u} j}|^2]=L_{\sf u} P$.
These information symbols will be delivered by $L_{\sf u}$ symbol extension, i.e., beamforming over $L_{\sf u}$ time slots.
In particular, let $\mathbf{W}_1 \in \mathbb{C}^{L_{\sf u} \times (L_{\sf u}-1)}$ be the submatrix consisting of the first through ($L_{\sf u}-1$)th column vectors of $\mathbf{\Omega}_{L_{\sf u}}$ and $\mathbf{w}_2 \in \mathbb{C}^{L_{\sf u} \times 1}$ be the $L_{\sf u}$th column of $\mathbf{\Omega}_{L_{\sf u}}$.
That is, $\mathbf{\Omega}_{L_{\sf u}} =\big[\mathbf{W}_1,\mathbf{w}_2\big]$.
The BS and the $j$th UL user set their length-$L_{\sf u}$ time-extended transmit signal vectors as
\begin{align}  \label{ex:input_theorem1}
\mathbf{x}_{ \sf d} \   =  \mathbf{W}_1 \mathbf{s}_{{\sf d} 1}, \ \mathbf{x}_{{\sf u}j}  = \mathbf{w}_2 s_{ {\sf u} j} \mbox{ for } j \in [1 : K_{\sf u}],
\end{align}
each of which satisfies the average power constraint $P$, i.e., $\mathbb{E}(\|\mathbf{x}_{ \sf d}\|^2)= L_{\sf u}P$ and $\mathbb{E}(\|\mathbf{x}_{{\sf u}j}\|^2)= L_{\sf u}P$ for $j \in [1 : K_{\sf u}]$.
Here, $\mathbf{W}_1$ is used as the transmit precoding matrix for sending $\mathbf{s}_{{\sf d} 1}$ and $\mathbf{w}_2$ is used as the transmit precoding vector for sending $s_{ {\sf u} j}$, which is the same for all $j\in[1:K_{\sf u}]$.
During signal transmission, the BS fixes its transmit mode, i.e., $\alpha(t)=1$ for all $t\in[1:L_{\sf u}]$. During signal reception, on the other hand, the BS sets its receive mode differently at each time, i.e., $\beta(t)=t$ for all $t\in[1:L_{\sf u}]$. Denote the above transmit mode vector and receive mode vector by $\bar{\alpha}_1$ and $\bar{\beta}_1$, respectively.

Then, from \eqref{eq:in_out_extended} and \eqref{ex:input_theorem1}, the length-$L_{\sf u}$ time-extended input--output relation is given by
\begin{align} 
\mathbf{y}_{ {\sf d} 1} & = h_{1}(1) \mathbf{W}_1 \mathbf{s}_{{\sf d} 1} + \mathbf{w}_2\sum_{j=1}^{K_{\sf u}} g_{1j} s_{ {\sf u} j}  + \mathbf{z}_{ {\sf d} 1}, \label{eq:received2}\\
\mathbf{y}_{ \sf u}& = \mathbf{R} \mathbf{s}_{\sf u} + \mathbf{z}_{\sf u} \label{eq:received3}
\end{align}
where $\mathbf{s}_{\sf u} = [s_{{\sf u}1}, \cdots, s_{{\sf u}K_{\sf u}}]^T$ and $\mathbf{R} = [ \mathbf{F}_1(\bar{\beta}_1)\mathbf{w}_2, \cdots, \mathbf{F}_{K_{\sf u}}(\bar{\beta}_1)\mathbf{w}_2 ]$.
Here, \eqref{eq:received3} holds from the fact that $\mathbf{H}_1(\bar{\alpha}_1)= h_1(1)\mathbf{I}_{L_{\sf u}}$.

For decoding its DL message, the first DL user multiplies $\mathbf{W}_1^H$ to $\mathbf{y}_{ {\sf d} 1}$, which is represented as
\begin{align} \label{eq:post_processing1}
\mathbf{W}_1^H \mathbf{y}_{ {\sf d} 1} = h_1(1)\mathbf{s}_{{\sf d} 1}+\mathbf{W}_1^H\mathbf{z}_{ {\sf d} 1} 
\end{align}
where the equality holds from \eqref{eq:property_IDFT}.
Then, the first DL user estimates its information symbols based on \eqref{eq:post_processing1}. 
%
Hence, the achievable DoF of the first DL user is
\begin{align}
d_{{\sf d}1} = 1-\frac{1}{L_{\sf u}} \notag.
\end{align}

Now consider decoding of $L_{\sf u}$ UL messages at the BS. The BS estimates its information symbols based on \eqref{eq:received3}.
From the definition of $\mathbf{R}$, $\mathbf{R}$ can be rewritten as $$\mathbf{R}=\operatorname{diag}(w_{21},\cdots,w_{2L_{\sf u}})[\mathbf{F}_1(\bar{\beta}_1),\cdots,\mathbf{F}_{K_{\sf u}}(\bar{\beta}_1)]$$ where $w_{2i}$ for $i \in [1:L_{\sf u}]$ is the $i$th element of $\mathbf{w}_2$ and thus $\operatorname{rank}(\mathbf{R}) = L_{\sf u}$ almost surely.
Therefore, from \eqref{eq:received3}, the achievable sum DoF of the $K_{\sf u}$ UL users is given by
\begin{align}
\sum_{j=1}^{K_{\sf u}} d_{{\sf u}j} = \frac{\operatorname{rank}(\mathbf{R})}{L_{\sf u}} = 1 \notag.
\end{align}

Consequently, the sum DoF of $2 - \frac{1}{L_{\sf u}}$ is achievable for no CSIT model, which completes the achievability proof of Theorem \ref{theorem:3}.

\subsection{Achievability for Theorem \ref{theorem:1} when $K_{\sf d}, K_{\sf u} \geq 1$} \label{proof:theorem2}

In this section, we show the achievability proof of Theorem \ref{theorem:1} when $K_{\sf d}, K_{\sf u} \geq 1$. 
For better understanding, we first illustrate the proposed scheme when $K_{\sf d} = K_{\sf u} = M_{\sf d} = M_{\sf u} = 2$ and then provide the achievability proof for the general case.

\subsubsection{Example case}
Consider the FD cellular network defined in Section \ref{sec:system} and assume that $K_{\sf d} = K_{\sf u} = M_{\sf d} = M_{\sf u} = 2$.
We now show that the transmitter of the BS sends two information symbols to each DL user and each UL user sends two information symbols to the receiver of the BS for four time slots ($n=4$).
As a result, the achievable sum DoF of the proposed scheme is given by two. For intuitive explanation, we skip the power constraint issue and some proof steps in this example case, which will be given in the next subsection.

Let $\mathbf{s}_{{\sf d}1}$, $\mathbf{s}_{{\sf d}2} \in \mathbb{C}^{2\times 1}$ be the information vectors sent to the first DL user and the second DL user 
and let $\mathbf{s}_{{\sf u}1}$, $\mathbf{s}_{{\sf u}2} \in \mathbb{C}^{2\times 1}$ be the information vectors sent by the first UL user and the second UL user.
Let $\mathbf{W}_3 \in \mathbb{C}^{4 \times 2}$ 
be the submatrix consisting of the first and the second columns of $\mathbf{\Omega}_{4}$ and
$\mathbf{W}_4 \in \mathbb{C}^{4 \times  2 }$ 
be the submatrix consisting of the third and the fourth columns of $\mathbf{\Omega}_{4}$.
Note that $\mathbf{\Omega}_4 = [\mathbf{W}_3,\mathbf{W}_4]$ and $\mathbf{W}_3^H \mathbf{W}_4 = \mathbf{0}_{2\times 2}$.
We set the transmit mode and the receive mode of the BS for 4 time slots, denoted by $\bar{\alpha}$ and $\bar{\beta}$ respectively, as $\bar{\alpha} = \bar{\beta} = [1, 2, 1, 2]^T$
and set the DL transmit precoding matrices as
\begin{align}
[ \mathbf{U}_1, \mathbf{U}_2 ] = 
\left [
\begin{array}{c}
\mathbf{W}_3^H \mathbf{H}_1(\bar{\alpha}) \\
\mathbf{W}_3^H \mathbf{H}_2(\bar{\alpha})
\end{array}
\right ]^{-1}.
\notag
\end{align}
Here we skip the proof of the existence of the above inverse matrix, which will be proved in the next subsection.
Then, the BS and the $j$th UL user construct their length-$4$ time-extended transmit signal vector as
\begin{align}
\mathbf{x}_{ \sf d} \  & = \mathbf{U}_1 \mathbf{s}_{ {\sf d} 1} + \mathbf{U}_2 \mathbf{s}_{ {\sf d} 2} , \ \mathbf{x}_{ {\sf u} j }  = \mathbf{W}_4 \mathbf{s}_{ {\sf u} j} \mbox{ for } j\in[1, 2] \notag
\end{align}

From \eqref{eq:in_out_extended}, the length-$4$ time-extended input--output relation is given by
\begin{align}
\mathbf{y}_{{\sf d} i} &= \mathbf{H}_i(\bar{\alpha}) ( \mathbf{U}_1 \mathbf{s}_{{\sf d}1}+ \mathbf{U}_{2}\mathbf{s}_{{\sf d}2}) + \sum_{j=1}^{2} g_{ij} \mathbf{W}_4 \mathbf{s}_{ {\sf u} j}  + \mathbf{z}_{ {\sf d} i} \mbox{ for } j\in[1, 2], \label{eq:inout_ex1} \\
\mathbf{y}_{ \sf u} &=[\mathbf{F}_1(\bar{\beta})\mathbf{V}, \mathbf{F}_{2}(\bar{\beta})\mathbf{W}_4 ][\mathbf{s}_{{\sf u}1}^T,\mathbf{s}_{{\sf u}2}^T]^T + \mathbf{z}_{\sf u} \label{eq:inout_ex2}
\end{align}
Then, the $i$th DL user estimates its information symbols by multiplying $\mathbf{W}_3^H$ to $\mathbf{y}_{{\sf d}i}$ in \eqref{eq:inout_ex1}.
From the definition of $\mathbf{U}_1$ and $\mathbf{U}_2$, 
\begin{align}
\mathbf{W}_3^H \mathbf{y}_{{\sf d} i} =  \mathbf{s}_{{\sf d}i} + \mathbf{W}_3^H \mathbf{z}_{{\sf d}i}  \mbox{ for } j\in[1, 2],
\end{align}
which shows that the $i$th DL user can obtain $\mathbf{s}_{{\sf d}i}$ almost surely.
The BS estimates its information symbols from \eqref{eq:inout_ex2}, showing that it can obtain $\mathbf{s}_{{\sf u}1}$ and $\mathbf{s}_{{\sf u}2}$ almost surely because
$[\mathbf{F}_1(\bar{\beta})\mathbf{V}, \mathbf{F}_{2}(\bar{\beta})\mathbf{V} ]$ is invertible almost surely, which will be proved in the next subsection.
Consequently, eight information symbols are delivered for four time slots and thus the achievable sum DoF of the proposed scheme is given by two.

\subsubsection{General proof}

Note that $L_{\sf d}, L_{\sf u} \geq 1$ from the assumption that $K_{\sf d}, K_{\sf u} \geq 1$.
In this case, the right-hand side of \eqref{eq:DoF} is given by
\begin{align}
\min \left \{ 2, \max \left( 1+ \frac{L_{\sf d}(L_{\sf u} - 1)}{L_{\sf u}} ,1+ \frac{L_{\sf u}(L_{\sf d} - 1)}{L_{\sf d}} \right )  \right \} .
\notag
\end{align}

In the following, we will show that the sum DoF of $\frac{n_{\sf d}}{L_{\sf u}} + \frac{n_{\sf u}}{L_{\sf d}}$ is achievable for all integer values $(n_{\sf d}, n_{\sf u})$ satisfying that
\begin{align} \label{eq:conditions}
n_{\sf d} &\in [1:L_{\sf u}],\nonumber\\
n_{\sf u} &\in [1:L_{\sf d}],\nonumber\\
n_{\sf d}+n_{\sf u} &\in[2: L_{\sf d} L_{\sf u}].
\end{align}
Notice that $(n_{\sf d}, n_{\sf u})=(L_{\sf u},\min( L_{\sf u}(L_{\sf d}-1) , L_{\sf d}))$ and $(n_{\sf d}, n_{\sf u})=(\min( L_{\sf d}(L_{\sf u}-1), L_{\sf u}),L_{\sf d})$ satisfy \eqref{eq:conditions}, which result in the sum DoFs of $\min\left( 2, 1+ \frac{L_{\sf u}(L_{\sf d} - 1)}{L_{\sf d}}\right)$ and $\min\left( 2, 1+ \frac{L_{\sf d}(L_{\sf u} - 1)}{L_{\sf u}}\right)$ respectively.
Then, the following relation holds:
\begin{align}
d_{\Sigma,  {\sf pCSIT}} & \geq \max \left \{ \min \left ( 2, 1+ \frac{L_{\sf d}(L_{\sf u} - 1)}{L_{\sf u}} \right ) , \min \left ( 2, 1+ \frac{L_{\sf u}(L_{\sf d} - 1)}{L_{\sf d}} \right )  \right \} \notag \\
												& = \min \left \{ 2, \max \left( 1+ \frac{L_{\sf d}(L_{\sf u} - 1)}{L_{\sf u}} ,1+ \frac{L_{\sf u}(L_{\sf d} - 1)}{L_{\sf d}} \right )  \right \}.
\end{align}
Therefore, in order to establish the achievablility of Theorem \ref{theorem:1}, it is enough to show that the sum DoF of $\frac{n_{\sf d}}{L_{\sf u}} + \frac{n_{\sf u}}{L_{\sf d}}$ is achievable for all integer values $(n_{\sf d}, n_{\sf u})$ satisfying that \eqref{eq:conditions}.  

From now on, assume that $(n_{\sf d}, n_{\sf u})$ satisfies \eqref{eq:conditions}. 
In the proof, the BS sends $n_{\sf d}$ information symbols to each of $L_{\sf d}$ DL users (out of $K_{\sf d}$ DL users) and each of $K_{\sf u}$ UL users sends $n_{\sf u}$ information symbols each to the BS for $L_{\sf d} L_{\sf u}$ time slots.

Let $\mathbf{s}_{ {\sf d} i} \in \mathbb{C}^{n_{\sf d}\times 1}$ be the information vector for the $i$th DL user, $i\in[1:L_{\sf d}]$, satisfying that $\mathbb{E}[\|\mathbf{s}_{{\sf d} i}\|^2] = n_{\sf d}  L_{\sf u}  P$.
Let $\mathbf{s}_{ {\sf u} j} \in \mathbb{C}^{n_{\sf u}\times 1}$ be the information vector for the $j$th UL user, where $j\in[1:K_{\sf u}]$, satisfying that $\mathbb{E}[\|\mathbf{s}_{{\sf u} j}\|^2]=n_{\sf u} L_{\sf d} L_{\sf u}P $.
These information symbols will be delivered by $L_{\sf d}L_{\sf u}$ symbol extension, i.e., beamforming over $L_{\sf d} L_{\sf u}$ time slots.
Let $\mathbf{U}_i \in \mathbb{C}^{ L_{\sf d} L_{\sf u} \times n_{\sf d} }$ be the transmit precoding matrix for sending $\mathbf{s}_{ {\sf d} i}$, where $i\in[1:L_{\sf d}]$, satisfying that $\sum_{i=1}^{L_{\sf d}} \|\mathbf{U}_{i}\|^2 = 1$ and
$\mathbf{V} \in \mathbb{C}^{ L_{\sf d} L_{\sf u} \times  n_{\sf u} }$ be the transmit precoding matrix for sending $\mathbf{s}_{ {\sf u} j}$, which is same for all $j\in[1:K_{\sf u}]$, 
satisfying that $\|\mathbf{V}\|^2 = 1$.
We will discuss designing of transmit precoding matrices of the BS and the UL users later.
The BS and the $j$th UL user set their length-$(L_{\sf d} L_{\sf u})$ time-extended transmit signal vector as
\begin{align}
\mathbf{x}_{ \sf d} \  & = \sum_{i=1}^{L_{\sf d}} \mathbf{U}_i \mathbf{s}_{ {\sf d} i} , \ \mathbf{x}_{ {\sf u} j }  = \mathbf{V} \mathbf{s}_{ {\sf u} j} \mbox{ for } j \in [1 : K_{\sf u}], \label{ex:input_theorem2}
\end{align}
each of which satisfies the average power constraint $P$, i.e., $\mathbb{E}(\|\mathbf{x}_{\sf d}\|^2) = L_{\sf d} L_{\sf u} P $ and $\mathbb{E}(\|\mathbf{x}_{{\sf u}j}\|^2) = L_{\sf d} L_{\sf u} P $ for $j \in [1:K_{\sf u}]$. 

During signal transmission and reception, the BS sets its transmit and receive mode differently at each time with cycle of $L_{\sf d}$ and $L_{\sf u}$ respectively, i.e., 
$\alpha(t) = (t-1)|L_{\sf d} + 1$ and $\beta(t) = (t-1)|L_{\sf u} + 1$ for $t\in[1:L_{\sf d} L_{\sf u}]$.
Denote the above transmit mode vector and receive mode vector by $\bar{\alpha}_2$ and $\bar{\beta}_2$, respectively.

Then, from \eqref{eq:in_out_extended} and \eqref{ex:input_theorem2}, the length-$(L_{\sf d} L_{\sf u})$ time-extended input--output relation is given by
\begin{align}
\mathbf{y}_{{\sf d} i} &=\mathbf{H}_i(\bar{\alpha}_2) [\mathbf{U}_1,\cdots, \mathbf{U}_{L_{\sf d}}] \mathbf{s}_{ {\sf d}} + \sum_{j=1}^{K_{\sf u}} g_{ij} \mathbf{V} \mathbf{s}_{ {\sf u} j}  + \mathbf{z}_{ {\sf d} i},\notag \\
\mathbf{y}_{ \sf u} &=[\mathbf{F}_1(\bar{\beta}_2)\mathbf{V}, \cdots, \mathbf{F}_{K_{\sf u}}(\bar{\beta}_2)\mathbf{V} ]\mathbf{s}_{\sf u} + \mathbf{z}_{\sf u} \label{eq:partial4}
\end{align}
where $\mathbf{s}_{\sf d} = [(\mathbf{s}_{{\sf d}1})^T,\cdots,(\mathbf{s}_{{\sf d} L_{\sf d}})^T]^T$ and $\mathbf{s}_{\sf u} = [(\mathbf{s}_{{\sf u}1})^T,\cdots,(\mathbf{s}_{{\sf u} K_{\sf u}})^T]^T$.

Now consider designing of the DL transmit precoding matrix $\mathbf{U}_{j}$ for $j \in [1:L_{\sf d}]$ and the UL transmit precoding matrix $\mathbf{V}$.
Let $\mathbf{W}_3 \in \mathbb{C}^{L_{\sf d} L_{\sf u} \times n_{\sf d}}$ 
be the submatrix consisting of the first through $n_{\sf d}$th columns of $\mathbf{\Omega}_{L_{\sf d} L_{\sf u}}$and
$\mathbf{W}_4 \in \mathbb{C}^{L_{\sf d} L_{\sf u} \times  n_{\sf u} }$ 
be the submatrix consisting of the ($n_{\sf d}+1$)th through ($n_{\sf d} + n_{\sf u}$)th columns of $\mathbf{\Omega}_{L_{\sf d} L_{\sf u}}$.
Let us define
\begin{align}
\mathbf{P} &= 
\left [
(\mathbf{W}_3^H \mathbf{H}_1(\bar{\alpha}_2) )^T, \cdots, (\mathbf{W}_3^H \mathbf{H}_{L_{\sf d}}(\bar{\alpha}_2) )^T
\right ]^T \in \mathbb{C}^{L_{\sf d} n_{\sf d} \times L_{\sf d} L_{\sf u} }, \notag \\
\mathbf{Q} & =  \left [ \mathbf{F}_1(\bar{\beta}_2) \mathbf{W}_4,\cdots, \mathbf{F}_{K_{\sf u}}(\bar{\beta}_2) \mathbf{W}_4 \right ]\in \mathbb{C}^{L_{\sf d} L_{\sf u} \times K_{\sf u} n_{\sf u} }  . \label{eq:partial5}
\end{align}
The following lemma is used for designing the transmit precoding matrices of the BS and the UL users.
\begin{lemma} \label{lemma:dl2}
$\operatorname{rank}(\mathbf{P}) = L_{\sf d} n_{\sf d}$ and
$\operatorname{rank}(\mathbf{Q}) \geq L_{\sf u} n_{\sf u}$ almost surely.
\end{lemma}
\begin{IEEEproof}
We refer to the Appendix for the proof.
\end{IEEEproof}

Now, we determine the transmit precoding matrices of the BS and the UL users as 
\begin{align}
[\mathbf{U}_1,\cdots, \mathbf{U}_{L_{\sf d}}] = \frac{\mathbf{P}^{\dagger}}{\|\mathbf{P}^{\dagger}\|}, \
\mathbf{V} = \frac{1}{\sqrt{n_{\sf u}}}\mathbf{W}_4 \label{eq:partial6}
\end{align}
where $\mathbf{P}^{\dagger}=\mathbf{P}^H (\mathbf{P} \mathbf{P}^H )^{-1}$ is the right inverse matrix of $\mathbf{P}$ satisfying that $\mathbf{P}\mathbf{P}^{\dagger} = \mathbf{I}_{L_{\sf d} n_{\sf d}}$, which exists almost surely from Lemma \ref{lemma:dl2}.

For decoding its DL message, the $i$th DL user multiplies $\mathbf{W}_3^H$ to $\mathbf{y}_{{\sf d}i}$. From \eqref{eq:partial4} and \eqref{eq:partial6},
\begin{align}
\mathbf{W}_3^H \mathbf{y}_{{\sf d} i} =  \frac{\mathbf{s}_{{\sf d}i}}{\|\mathbf{P}^{\dagger}\| } + \mathbf{W}_3^H \mathbf{z}_{{\sf d}i} \label{eq:partial3}
\end{align}
where the equality holds from the definition of $\mathbf{P}$ in \eqref{eq:partial5} and the property of the IDFT matrix in \eqref{eq:property_IDFT}.
%
Then, the $i$th DL user estimates its information symbols based on \eqref{eq:partial3}. 
Hence, the achievable sum DoF of the $L_{\sf d}$ DL users is given by
\begin{align}
\sum_{i=1}^{L_{\sf d}} d_{{\sf d}i} = \frac{n_{\sf d}}{L_{\sf u}} \notag.
\end{align}

Now consider decoding of the UL messages at the BS. 
From \eqref{eq:partial4} and \eqref{eq:partial6}, the received signal of the BS is given by
\begin{align}
\mathbf{y}_{\sf u} = \mathbf{Q}\mathbf{s}_{\sf u} + \mathbf{z}_{\sf u}. \label{eq:inout_general}
\end{align}
Then, the BS estimates its information symbols based on \eqref{eq:inout_general}, provided that 
the achievable sum DoF of the $K_{\sf u}$ UL users is given by
\begin{align}
\sum_{i=1}^{K_{\sf u}} d_{{\sf u}i} = \frac{\operatorname{rank}(\mathbf{Q})}{L_{\sf d} L_{\sf u}} \geq \frac{n_{\sf u}}{L_{\sf d}} \notag
\end{align}
where the inequality follows from Lemma \ref{lemma:dl2}.

Consequently, the sum DoF of $\frac{n_{\sf d}}{L_{\sf u}} + \frac{n_{\sf u}}{L_{\sf d}}$ is achievable for all $n_{\sf d} \in [1:L_{\sf u}]$ and $n_{\sf u} \in [1:L_{\sf d}]$ satisfying $n_{\sf d}+n_{\sf u} \in[2: L_{\sf d} L_{\sf u}]$,
which completes the achievability of Theorem \ref{theorem:1}.

\section{Converse} \label{sec:converse}

In this section, we establish the converse of Theorem \ref{theorem:3}. 
When $K_{\sf d}=0$ or $K_{\sf u}=0$, the right-hand side of \eqref{eq:DoF11} in Theorem \ref{theorem:3} is given by
\begin{align}
\min \big \{ \max(K_{\sf d}, K_{\sf u}), 1 \big \} \ \mbox{ when } K_{\sf d} = 0 \mbox{ or } K_{\sf u} = 0, \notag 
\end{align}
which holds from the sum DoF of broadcast channels and multiple-access channels \cite{Davoodi:16, Tse:04}.

\begin{figure}[!t] 
\includegraphics[width=2.5in]{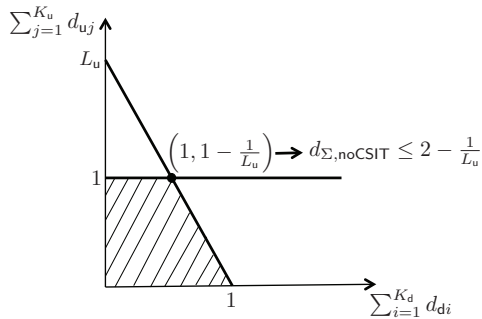}
\centering 
\caption{Feasible region of $(\sum_{i=1}^{K_{\sf d}}d_{{\sf d}i},\sum_{j=1}^{K_{\sf u}}d_{{\sf u}j})$.} \label{fig:feasible}
\end{figure}

Now, we show the converse proof of Theorem \ref{theorem:3} when $K_{\sf d}, K_{\sf u} \geq 1$ for the rest of this section.
We first introduce the following key lemma.
\begin{lemma} \label{lemma:1}
For the FD cellular network with no CSIT, any achievable DoF tuple must satisfy the following
inequality:
\begin{align} \label{eq:DoF_upper_bound}
\sum_{i=1}^{K_{\sf d}} d_{{\sf d} i} + \frac{1}{\min(K_{\sf u},M_{\sf u})}\sum_{j=1}^{K_{\sf u}} d_{{\sf u} j} \leq 1. 
\end{align}
\end{lemma}
\begin{IEEEproof}
We refer to Section \ref{sec:prooflem1} for the proof.
\end{IEEEproof}
For notational convenience, let $d_{\sf d}=\sum_{i=1}^{K_{\sf d}}d_{{\sf d}i}$, $d_{\sf u}=\sum_{j=1}^{K_{\sf u}}d_{{\sf u}j}$, and $L_{\sf u}=\min(K_{\sf u},M_{\sf u})$.
Then \eqref{eq:DoF_upper_bound} is rewritten as
\begin{align} \label{eq:DoF_upper_bound2}
d_{{\sf d}} + \frac{1}{L_{\sf u}}d_{{\sf u}} \leq 1. 
\end{align}
Trivially, from the sum DoF of the multiple-access channel \cite{Tse:04}, we have
$d_{{\sf u}} \leq 1$. 
Therefore, any achievable $(\sum_{i=1}^{K_{\sf d}}d_{{\sf d}i},\sum_{j=1}^{K_{\sf u}}d_{{\sf u}j})$ pair should be located inside the shaded region in Fig. \ref{fig:feasible}.
In conclusion, $d_{\Sigma,{\sf noCSIT}}\leq 2-\frac{1}{L_{\sf u}}$, which completes the proof of Theorem \ref{theorem:3}. For the rest of this section, we prove Lemma \ref{lemma:1}.

\subsection{Proof of Lemma \ref{lemma:1}} \label{sec:prooflem1}

\begin{figure}[!t] 
\includegraphics[width=2.5in]{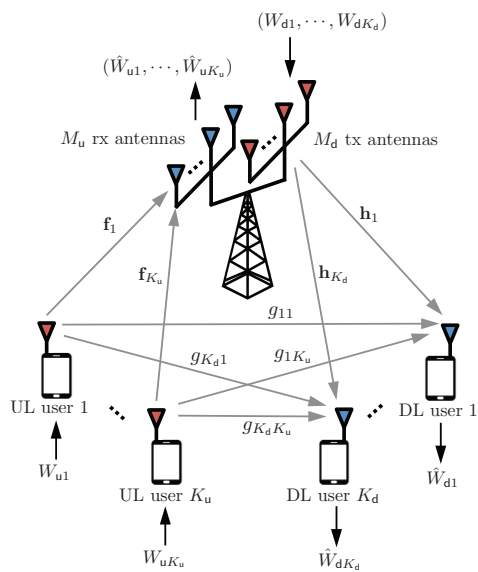}
\centering 
\caption{Extended networks having $M_{\sf d}$ tx antennas and $M_{\sf u}$ rx antennas at the BS.} \label{fig:extended}
\end{figure}

\subsubsection{Extended networks}
To prove Lemma \ref{lemma:1}, we first introduce the extended network in Fig. \ref{fig:extended} consisting of $M_{\sf d} $ and $M_{\sf u} $ conventional antennas at the transmitter and the receiver of the BS, instead of reconfigurable antennas each of which can choose a single transmit and receive mode from $M_{\sf d} $ and $M_{\sf u} $ preset modes.
Obviously, the achievable DoF region of the original network is included in that of the extended network.

More specifically, the received signal of the $i$th DL user at time $t$ and the received signal vector of the  BS at time $t$ are given respectively by
\begin{align}
y_{ {\sf d} i}(t) & = \mathbf{h}_i  \mathbf{x}_{\sf d}(t) + \sum_{j=1}^{K_{\sf u}} g_{ij} x_{ {\sf u} j}(t) + z_{{\sf d} i}(t), \notag \\
\mathbf{y}_{\sf u}(t) & = \sum_{j=1}^{K_{\sf u}} \mathbf{f}_j x_{ {\sf u} j}(t) + \mathbf{z}_{\sf u}(t) \label{eq:inout}
\end{align}
where 
$\mathbf{x}_{\sf d}(t) \in \mathbb{C}^{M_{\sf d} \times 1}$ is the transmit signal vector of the BS at time $t$, $x_{{\sf u} j}(t) \in \mathbb{C}$ is the transmit signal of the $j$th UL user at time $t$,
$\mathbf{h}_i \in \mathbb{C}^{1 \times M_{\sf d}}$ is the channel vector from the transmitter of the BS to the $i$th DL user, 
$g_{ij} \in \mathbb{C}$ is the channel from the $j$th UL user to the $i$th DL user, and
$\mathbf{f}_j \in \mathbb{C}^{M_{\sf u}  \times 1}$ is the channel vector from the $j$th UL user to the receiver of the BS.
The elements in additive noises $z_{{\sf d} i}(t) \in \mathbb{C}$ and $\mathbf{z}_u(t)\in \mathbb{C}^{M_{\sf u}\times 1}$ are i.i.d. drawn from $\mathcal{CN}(0,1)$.
The BS and each UL user should satisfy the average power constraint $P$, i.e., $\mathbb{E} \left[ \|\mathbf{x}_d(t)\|^2 \right ]   \leq P$ and $\mathbb{E} \left[ \|x_{{\sf u} j}(t)\|^2 \right ]  \leq P$ for all $j \in [1 : K_{\sf u}]$.
In the same manner as in Section \ref{sec:system1}, we assume that all channel coefficients are i.i.d. drawn from a continuous distribution and CSIT is not available at the BS and each UL user (no CSIT model). 
Then we can define the sum DoF of the extended model in the same manner as in Section \ref{sec:DoF}.

From \eqref{eq:inout}, the length-$n$ time-extended input--output relation is given by
\begin{align} 
\mathbf{y}_{{\sf d}i} & = \mathbf{H}_{i}\mathbf{x}_{\sf d} + \sum_{j=1}^{K_{\sf u}} g_{ij} \mathbf{x}_{{\sf u}j}  + \mathbf{z}_{{\sf d}i}, \notag \\
\mathbf{y}_{\sf u} & = \sum_{j=1}^{K_{\sf u}} \mathbf{F}_{j} \mathbf{x}_{{\sf u}j} + \mathbf{z}_{\sf u} \notag
\end{align}
where
\begin{align*}
\mathbf{H}_i & = \mathbf{I}_n \otimes \mathbf{h}_i, \ \mathbf{F}_j  = \mathbf{I}_n \otimes \mathbf{f}_j,  \\
\mathbf{y}_{{\sf d}i} & = \left[y_{{\sf d}i}(1), \cdots, y_{{\sf d}i}(n) \right]^T, \ \mathbf{y}_{\sf u}  = \left[\mathbf{y}_{\sf u}(1)^T, \cdots, \mathbf{y}_{\sf u}(n)^T \right]^T,\\
\mathbf{x}_{\sf d} 	& = \left[\mathbf{x}_{\sf d}(1)^T,\cdots,\mathbf{x}_{\sf d}(n)^T\right]^T, \ \mathbf{x}_{{\sf u}i} = \left[x_{{\sf u}i}(1),\cdots,x_{{\sf u}i}(n)\right]^T, \\
\mathbf{z}_{{\sf d}i}& = \left[z_{{\sf d}i}(1), \cdots, z_{{\sf d}i}(n)\right]^T, \ \mathbf{z}_{{\sf u}} = \left[\mathbf{z}_{{\sf u}}(1)^T, \cdots, \mathbf{z}_{{\sf u}}(n)^T\right]^T.
\end{align*}

\subsubsection{DoF upper bound}

We will prove that any DoF tuple achievable for the extended network in Fig. \ref{fig:extended} satisfies \eqref{eq:DoF_upper_bound2}.
Let $\mathbf{F} = [\mathbf{f}_1, \cdots, \mathbf{f}_{K_{\sf u}}] \in \mathbb{C}^{M_{\sf u}  \times K_{\sf u}}$ be the compound channel matrix from $K_{\sf u}$ UL user to the receiver of the BS. In order to establish \eqref{eq:DoF_upper_bound2}, we decompose $\mathbf{y}_{{\sf u}}(t)$, $\mathbf{z}_{\sf u}(t)$, and  $\mathbf{F}$ as follows:
\begin{itemize}
\item Decompose $\mathbf{y}_{\sf u}(t)$ into $\mathbf{y}_{{\sf u} \alpha}(t) \in \mathbb{C}^{L_{\sf u} \times 1}$ and $\mathbf{y}_{{\sf u} \beta}(t) \in \mathbb{C}^{(M_{\sf u} -L_{\sf u})\times 1}$
such that $$\mathbf{y}_{\sf u}(t) = \left [ \mathbf{y}_{{\sf u} \alpha}(t)^T,  \mathbf{y}_{{\sf u} \beta}(t)^T  \right]^T$$
and let $\mathbf{y}_{{\sf u}\alpha}=[\mathbf{y}_{{\sf u}\alpha}(1)^T,\cdots,\mathbf{y}_{{\sf u}\alpha}(n)^T]^T$ 
and $\mathbf{y}_{{\sf u}\beta}=[\mathbf{y}_{{\sf u}\beta}(1)^T,\cdots,\mathbf{y}_{{\sf u}\beta}(n)^T]^T$.
\item Decompose $\mathbf{z}_{\sf u}(t)$ into
$\mathbf{z}_{{\sf u} \alpha}(t) \in \mathbb{C}^{L_{\sf u}\times 1}$ and $\mathbf{z}_{{\sf u} \beta}(t) \in \mathbb{C}^{(M_{\sf u}-L_{\sf u})\times 1}$
such that $$\mathbf{z}_{\sf u}(t) = \left [ \mathbf{z}_{{\sf u} \alpha}(t)^T , \mathbf{z}_{{\sf u} \beta}(t)^T  \right]^T$$
and let $\mathbf{z}_{{\sf u}\alpha}=[\mathbf{z}_{{\sf u}\alpha}(1)^T,\cdots,\mathbf{z}_{{\sf u}\alpha}(n)^T]^T$ 
and $\mathbf{z}_{{\sf u}\beta}=[\mathbf{z}_{{\sf u}\beta}(1)^T,\cdots,\mathbf{z}_{{\sf u}\beta}(n)^T]^T$.
\item Decompose $\mathbf{F}$ 
into $\mathbf{F}_{\alpha} \in \mathbb{C}^{L_{\sf u} \times K_{\sf u}}$,
and $\mathbf{F}_{\beta} \in \mathbb{C}^{(M_{\sf u} -L_{\sf u}) \times K_{\sf u}}$ 
such that $\mathbf{F} =\left [\mathbf{F}_{\alpha}^T, \mathbf{F}_{\beta}^T\right ]^T$.
\end{itemize}
%
%
Furthermore, we define 
\begin{align} \label{eq:tilde_y}
\tilde{\mathbf{y}}_{{\sf u}\alpha}(t) = \mathbf{y}_{{\sf u}\alpha}(t) + \mathbf{T} \mathbf{x}_{\sf d}(t)
\end{align}
and $\tilde{\mathbf{y}}_{{\sf u}\alpha}=[\tilde{\mathbf{y}}_{{\sf u}\alpha}(1)^T,\cdots,\tilde{\mathbf{y}}_{{\sf u}\alpha}(n)^T]^T$, where all coefficients in $\mathbf{T} \in \mathbb{C}^{L_{\sf u} \times M_{\sf d} }$ are i.i.d. drawn from the distribution of the channel coefficients.
For convenience, let us denote the set of all channel coefficients, the set of DL messages, and the set of UL messages by 
\begin{align}
\mathcal{H} & = \big \{ \{\mathbf{h}_i\}_{i\in[1:K_{\sf d}]}, \{g_{ij}\}_{i\in[1:K_{\sf d}], j\in[1:K_{\sf u}]}, \{\mathbf{f}_j\}_{ j\in[1:K_{\sf u}]}, \mathbf{T} \big \} \notag \\
\mathcal{W}_{\sf d} & = (W_{ {\sf d} 1},\cdots, W_{ {\sf d} K_{\sf d} }), \ \mathcal{W}_{\sf u}  = (W_{ {\sf u} 1},\cdots, W_{ {\sf u} K_{\sf u} }). \notag
\end{align}

We are now ready to prove \eqref{eq:DoF_upper_bound} under the extended network.
From Fano's inequality, we have
\begin{align}
R_{ {\sf d} i} - \epsilon_{n} & \leq \frac{1}{n}\mathcal{I}\left (W_{ {\sf d} i};\mathbf{y}_{{\sf d} i}| \mathcal{H}, W_{ {\sf d} 1},\cdots, W_{ {\sf d} i-1} \right) \notag \\
											 & = \frac{1}{n}\mathcal{I}\left (W_{ {\sf d} i};\mathbf{y}_{{\sf d} 1}| \mathcal{H}, W_{ {\sf d} 1},\cdots, W_{ {\sf d} i-1}\right ) \notag 
\end{align}
where $\epsilon_n\geq 0$ converges to zero as $n$ increases.
Here the equality holds from the fact that the conditional probability distribution of $\mathbf{y}_{{\sf d}i}$ is the same for all $i\in[1:K_{\sf d}]$ when $(\mathcal{H}, W_{ {\sf d} 1},\cdots, W_{ {\sf d} i})$ is given.
Subsequently, 
\begin{align} 
\sum_{i=1}^{K_{\sf d}} R_{ {\sf d} i} -K_{\sf d}\epsilon_n &  \leq \frac{1}{n}\sum_{i=1}^{K_{\sf d}} \mathcal{I} \left (W_{ {\sf d} i}; \mathbf{y}_{{\sf d} 1}| \mathcal{H}, W_{ {\sf d} 1},\cdots, W_{ {\sf d} i-1} \right ) \notag \\
											& = \frac{1}{n}\mathcal{I} \left (\mathcal{W}_{ {\sf d} }; \mathbf{y}_{{\sf d} 1}| \mathcal{H}\right) \notag \\
											&  =   \frac{1}{n}h \left (\mathbf{y}_{{\sf d} 1}| \mathcal{H} \right ) - \frac{1}{n}h \left (\mathbf{y}_{{\sf d} 1}| \mathcal{H},\mathcal{W}_{ {\sf d} } \right ) \notag \\
											&  \leq   \log P - \frac{1}{n}h\left (\mathbf{y}_{{\sf d} 1}| \mathcal{H},\mathcal{W}_{ {\sf d} } \right )  +o(\log P) \label{eq:receiver1_converse}
\end{align}
where the last inequality holds since  $h \left (\mathbf{y}_{{\sf d} 1}| \mathcal{H} \right )\leq  n(\log P+o(\log P))$.

From Fano's inequality, we have 
\begin{align}
R_{ {\sf u} j} - \epsilon_n & \leq \frac{1}{n}\mathcal{I} \left (W_{ {\sf u} j};\mathbf{y}_{\sf u}| \mathcal{H}, W_{ {\sf u} 1},\cdots, W_{ {\sf u} j-1} \right ) \notag 
\end{align}
yielding that 
\begin{align} \label{eq:receiver2_converse}
\sum_{j=1}^{K_{\sf u}}  R_{ {\sf u} j} - K_{\sf u} \epsilon_n& \leq \frac{1}{n}\sum_{j=1}^{K_{\sf u}} \mathcal{I}\left (W_{ {\sf u} j}; \mathbf{y}_{\sf u}| \mathcal{H}, W_{ {\sf u} 1}, \cdots, W_{ {\sf u} j-1} \right ) \notag \\
											 &  \overset{(a)}{=}   \frac{1}{n}\mathcal{I} \left (\mathcal{W}_{ {\sf u} }; \mathbf{y}_{\sf u}| \mathcal{H}, \mathcal{W}_{ {\sf d} } \right ) \notag \\
											 &   =    \frac{1}{n}h \left (\mathbf{y}_{\sf u}| \mathcal{H}, \mathcal{W}_{ {\sf d} } \right) - \frac{1}{n}h \left (\mathbf{y}_{\sf u}| \mathcal{H},\mathcal{W}_{ {\sf d} },\mathcal{W}_{ {\sf u} } \right ) \notag \\
											 & \overset{(b)}{\leq}  \frac{1}{n}h \left ( \mathbf{y}_{\sf u}| \mathcal{H}, \mathcal{W}_{ {\sf d} } \right )  \notag\\
											 & \overset{(c)}{\leq}  \frac{1}{n}h \left ( \tilde{\mathbf{y}}_{{\sf u}\alpha}| \mathcal{H}, \mathcal{W}_{ {\sf d} } \right )  
\end{align}
where $(a)$ holds from chain rules for mutual information and the fact that $\mathcal{W}_{\sf d}$ is independent of $(\mathcal{W}_{\sf u},\mathbf{y}_{\sf u})$, 
$(b)$ holds from the fact that the differential entropy of white Gaussian noise is non-negative
and $(c)$ holds from 
\begin{align*}
		& h \left (\mathbf{y}_{\sf u}| \mathcal{H}, \mathcal{W}_{\sf d} \right )  \notag \\
		& = \sum\limits_{t=1}^{n} h \left (\mathbf{y}_{\sf u}(t)| \mathcal{H}, \mathcal{W}_{\sf d}, \mathbf{y}_{\sf u}(1),\cdots, \mathbf{y}_{\sf u}(t-1) \right)  \notag \\
		& = \sum\limits_{t=1}^{n}  h \left (\mathbf{y}_{{\sf u} \alpha}(t)| \mathcal{H}, \mathcal{W}_{ {\sf d} }, \mathbf{y}_{\sf u}(1),\cdots, \mathbf{y}_{\sf u}(t-1)\right) \notag \\ 
		& \ \   +\sum\limits_{t=1}^{n}  h \left (\mathbf{y}_{{\sf u} \beta}(t)| \mathcal{H}, \mathcal{W}_{ {\sf d} }, \mathbf{y}_{\sf u}(1),\cdots, \mathbf{y}_{\sf u}(t-1), \mathbf{y}_{{\sf u} \alpha}(t) \right )   \notag\\
		& \overset{(a)}{\leq} \sum\limits_{t=1}^{n}  h \left (\mathbf{y}_{{\sf u} \alpha}(t)| \mathcal{H}, \mathcal{W}_{ {\sf d} }, \mathbf{y}_{\sf u}(1),\cdots, \mathbf{y}_{\sf u}(t-1)\right)  
		+ n \cdot o(\log P ) \notag\\
		& \overset{(b)}{=}  \sum\limits_{t=1}^{n} h \left (\tilde{\mathbf{y}}_{{\sf u}\alpha}(t)| \mathcal{H}, \mathcal{W}_{ {\sf d} }, \mathbf{y}_{\sf u}(1),\cdots, \mathbf{y}_{\sf u}(t-1), \tilde{\mathbf{y}}_{{\sf u}\alpha}(1) ,\cdots, \tilde{\mathbf{y}}_{{\sf u}\alpha}(t-1)\right ) + n \cdot o(\log(P)) \notag\\
		& \overset{(c)}{\leq}   \sum\limits_{t=1}^{n} h \left (\tilde{\mathbf{y}}_{{\sf u}\alpha}(t)| \mathcal{H}, \mathcal{W}_{ {\sf d} }, \tilde{\mathbf{y}}_{{\sf u}\alpha}(1) ,\cdots, \tilde{\mathbf{y}}_{{\sf u}\alpha}(t-1) \right) + n \cdot o(\log(P)) \notag\\
		 & = h \left (\tilde{\mathbf{y}}_{{\sf u}\alpha}| \mathcal{H}, \mathcal{W}_{ {\sf d} } \right ) + n \cdot o(\log(P)) \notag 
\end{align*}
where $(a)$ holds from the fact that
if $K_{\sf u} < M_{\sf u}$, then 
$
\mathbf{y}_{{\sf u} \beta}(t)  
= \mathbf{F}_{\beta}
\mathbf{F}_{\alpha}^{-1} 
(\mathbf{y}_{{\sf u} \alpha}(t)-\mathbf{z}_{{\sf u} \alpha}(t))
+ 
\mathbf{z}_{{\sf u} \beta}(t)
$
and otherwise, $\mathbf{y}_{{\sf u} \beta}(t)$ does not exist from its definition, $(b)$ holds from the fact that $\tilde{\mathbf{y}}_{{\sf u}\alpha}(t)$ is a function of $\{\mathcal{H}, \mathcal{W}_{ {\sf d} }, \mathbf{y}_{\sf u}(1), \cdots, \mathbf{y}_{\sf u}(t-1), \mathbf{y}_{{\sf u}\alpha}(t)\}$  for $t \in [1:n]$ from the definition in \eqref{eq:tilde_y}, and $(c)$ holds from the fact that conditioning reduces differential entropy.

Let $\tilde{y}_{{\sf u}\alpha i}(t) \in \mathbb{C}$ for $i \in [1: L_{\sf u}]$ be the $i$th element of $\tilde{\mathbf{y}}_{{\sf u}\alpha}(t)$ and
$\tilde{\mathbf{y}}_{{\sf u}\alpha i} = [\tilde{y}_{{\sf u}\alpha i}(1),\cdots,\tilde{y}_{{\sf u}\alpha i}(n)]^T$.
%
From the definition of $\tilde{\mathbf{y}}_{{\sf u}\alpha}(t)$ in \eqref{eq:tilde_y}, the conditional probability distribution of $\mathbf{y}_{{\sf d} 1}$ is identical with that of $\tilde{\mathbf{y}}_{{\sf u}\alpha i}$ for all $i \in [ 1 : L_{\sf u} ]$ when $(\mathcal{H}, \mathcal{W}_{ {\sf d} })$ is given.
Consequently, from \eqref{eq:receiver2_converse} 
\begin{align} \label{eq:convese22}
\sum_{j=1}^{K_{\sf u}}  R_{ {\sf u} j} - K_{\sf u} \epsilon_n& \leq\frac{1}{n}h \left ( \tilde{\mathbf{y}}_{{\sf u}\alpha}| \mathcal{H}, \mathcal{W}_{ {\sf d} } \right ) +  o(\log P) \nonumber\\
&\leq\frac{1}{n}\sum_{i=1}^{L_{\sf u}}h \left ( \tilde{\mathbf{y}}_{{\sf u}\alpha i}| \mathcal{H}, \mathcal{W}_{ {\sf d} } \right ) +  o(\log P)  \nonumber\\
&=\frac{1}{n}L_{\sf u}h \left ( \mathbf{y}_{{\sf d}1}| \mathcal{H}, \mathcal{W}_{ {\sf d} } \right ) +  o(\log P)  
\end{align}
where the second inequality holds from the fact that conditioning reduces differential entropy.

Then, multiplying  $\frac{1}{L_{\sf u}}$ to \eqref{eq:convese22} and adding it to \eqref{eq:receiver1_converse}, we have
\begin{align} \label{eq:converse3}
\sum_{i=1}^{K_{\sf d}} R_{ {\sf d} i} +\frac{1}{L_{\sf u}}\sum_{j=1}^{K_{\sf u}}  R_{ {\sf u} j} \leq \log P + o(\log P) +\left(K_{\sf d}+\frac{1}{L_{\sf u}}K_{\sf u}\right) \epsilon_n.
\end{align}
By dividing both hand sides of \eqref{eq:converse3} by $\log P$ and letting $n$ and $P$ to infinity, we have
\begin{align}
\sum_{i=1}^{K_{\sf d}} d_{{\sf d} i} + \frac{1}{L_{\sf u}}\sum_{j=1}^{K_{\sf u}} d_{{\sf u} j} \leq 1 \notag
\end{align}
where $\epsilon_n$ converges to zero as $n$ increases, which completes the proof of Lemma \ref{lemma:1}.

\section{Sum Rate Comparison} \label{sec:sum_rates}

In this section, we numerically demonstrate the sum rate improvement of the proposed schemes (FD systems) by comparing with HD systems at the finite SNR regime.
For comprehensive comparison, we also consider multicell environment and the impacts of residual self-interference due to imperfect self-interference suppression and user scheduling.

\subsection{Single-Cell Case} \label{subsec:sum_rates}

In this subsection, we compare the average sum rates of the proposed schemes with those of the conventional HD systems for both no CSIT and partial CSIT models.
To evaluate the sum rates of HD systems, we assume that the BS operates in TDD and the half fraction of time resource is allocated for DL transmission and the rest half fraction is allocated for UL transmission and further assume that all UL users simultaneously transmit to the BS to maximize the UL sum rate and, on the other hand, the BS transmits to a single DL user to maximize the DL sum rate \cite{Tse:05}.
%
The only difference between no CSIT and partial CSIT models in HD systems is the fact that the BS can choose the transmit mode of the reconfigurable antenna and the serving DL user in order to maximize the DL sum rate for the partial CSIT model. For no CSIT model, on the other hand, the BS randomly chooses its transmit mode and serving DL user. 

In order to reflect the sum rate degradation of the proposed schemes due to imperfect self-interference suppression, we assume residual self-interference at the BS.  
In \cite{Bharadia:13}, the authors propose novel analog and digital SIC techniques with SIC capability of 110 dB and show that residual self-interference can be reduced almost to the same level as the noise power when the average transmit power is around 20 dBm, which corresponds to the transmit power used in commercial communication systems such as WiFi and LTE small cell.
From \cite{Bharadia:13}, we assume that the residual self-interference power is assumed to be the same as the noise power and regarded it as noise in simulation.

\begin{figure}[!t]
\includegraphics[width=3.2in]{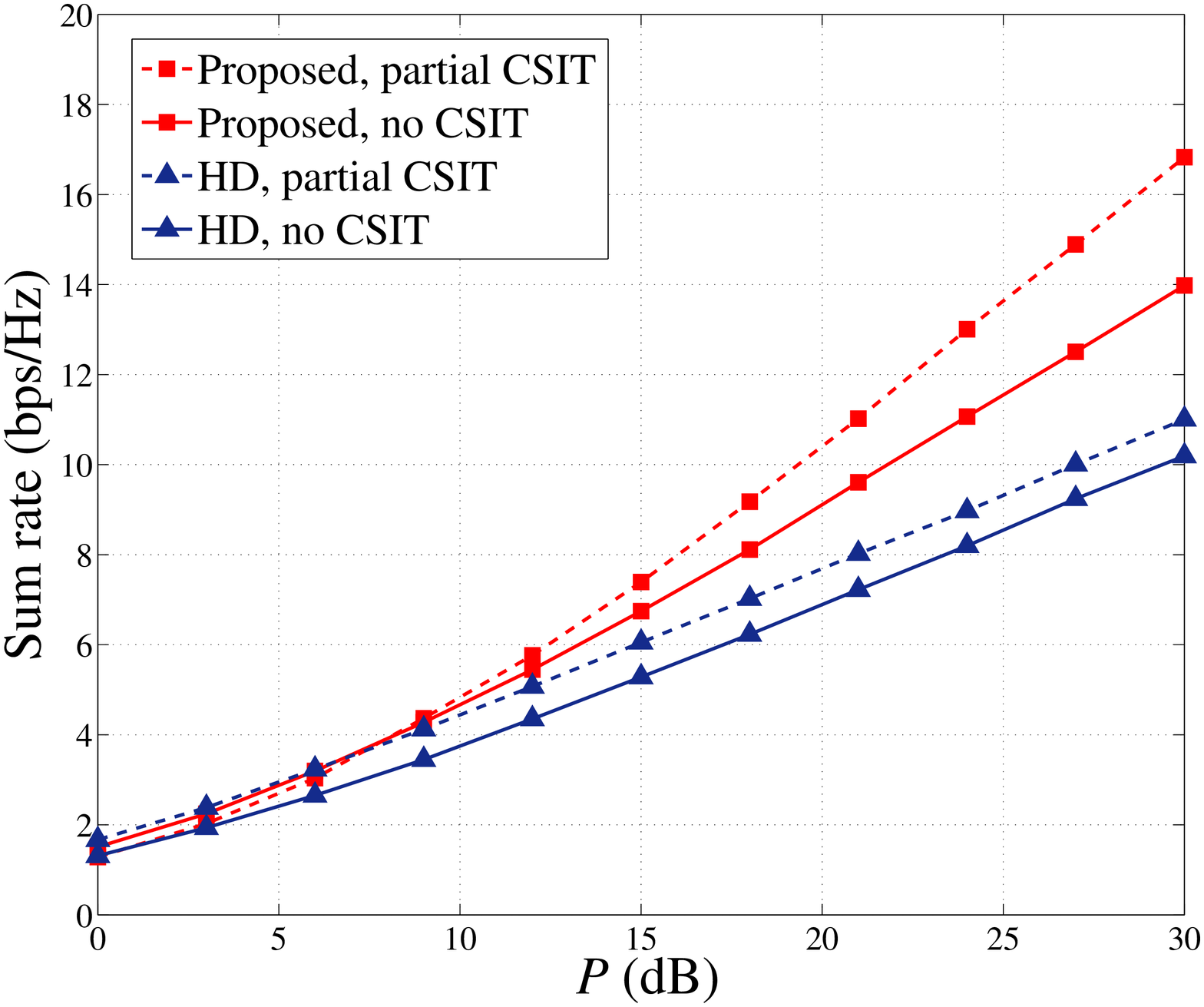}
\centering
\caption{Average sum rate in single-cell environment when $K_{\sf d} = K_{\sf u} = M_{\sf d} = M_{\sf u} = 2$.}
\label{rate_symmetric}
\end{figure}

Fig. \ref{rate_symmetric} plots the average sum rates of the proposed schemes and the conventional HD systems with respect to $P$ when $K_{\sf d} = K_{\sf u} = M_{\sf d} = M_{\sf u} = 2$.
All channel coefficients are assumed to be i.i.d. drawn from the circularly symmetric complex Gaussian distribution, i.e., $\mathcal{CN}(0,1)$.
As seen in the figure, the proposed schemes gradually outperform the conventional HD systems and, moreover, the sum rate gaps increase as SNR increases. Hence, the DoF gains achievable by the proposed schemes actually yield the sum rate gains at the finite and operational SNR regime.

\subsection{Multicell Case}

\begin{figure}[!t]
\includegraphics[width=3.2in]{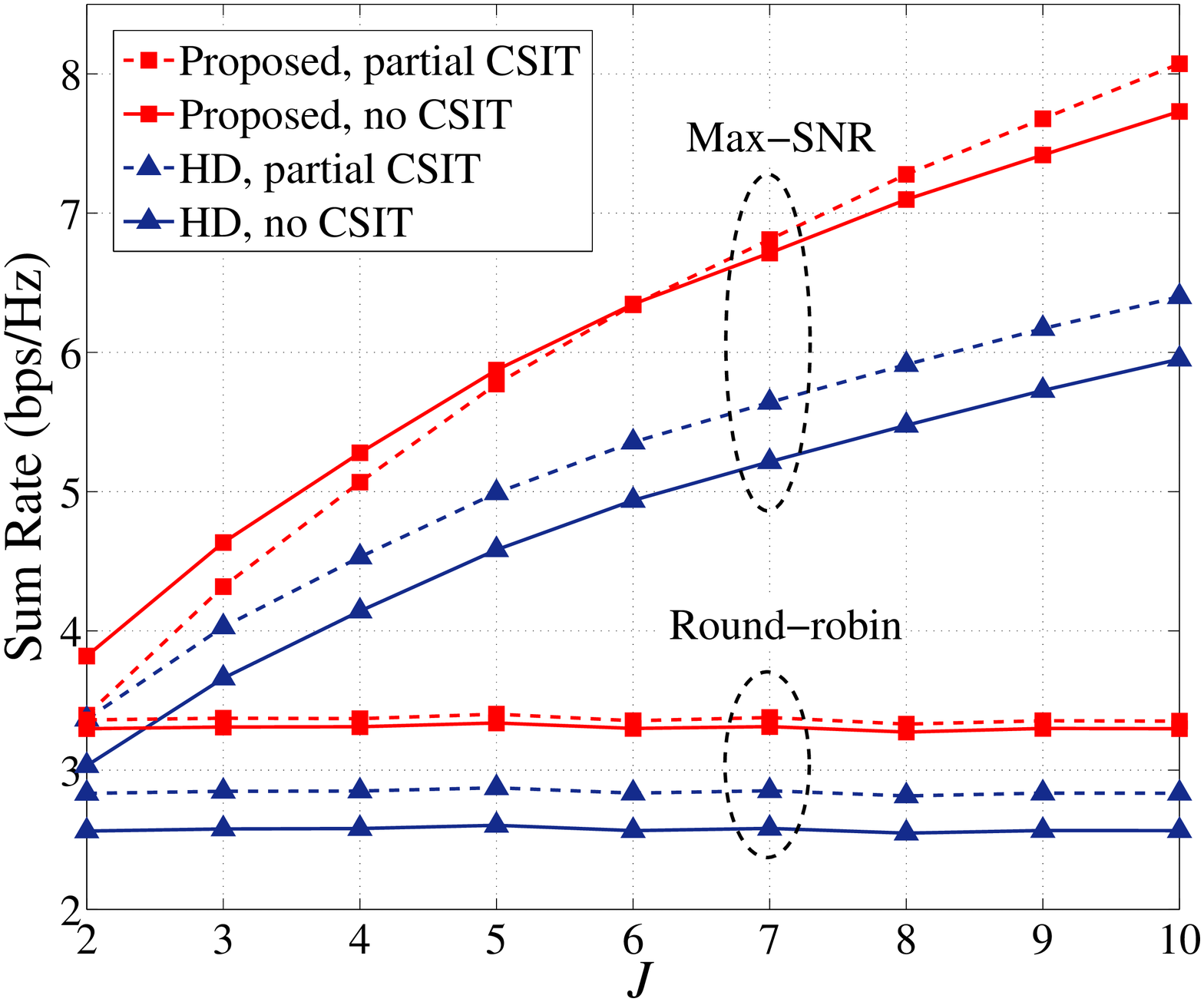}
\centering
\caption{Average sum rates in multicell environment with respect to $J$ when $K_{\sf d} = K_{\sf u} = M_{\sf d} = M_{\sf u} = 2$, $\alpha = 3$, and $P_{\sf ref}= 10$ dB.}
\label{rate_symmetric2}
\end{figure}

In this subsection, we consider multicell environment and compare the sum rates of the proposed schemes with those of the conventional HD systems.
Specifically, we adopt a wrap-around configuration of 7 hexagonal cells in which each BS is located in the center of each cell and the maximum distance from the center within each cell is given by one.
We evaluate the sum rates at the center cell by treating inter-cell interference arisen from other six cells as noise.
We assume that $J$ DL users and $J$ UL users are distributed uniformly at random within the area of each cell.
A simplified path-loss channel model is used with path-loss exponent $\alpha$ and for convenience we denote the average received SNR at the maximum distance of one by $P_{\sf ref}$ \cite{Goldsmith:05}.
For the center cell, the channel coefficient from the $n$th preset mode of the transmit antenna of the BS to the $i$th DL user, 
the channel coefficient from the $j$th UL user to the $m$th preset mode of the receive antenna of the BS,
the channel coefficient from the $j$th UL user to the $i$th DL user at the center cell are given respectively by
\begin{align}
h_i(n) = \frac{\hat{h}_i(n)}{d_{{\sf dl}, i}^{\alpha/2}}, \
f_j(m) = \frac{\hat{f}_j(m)}{ d_{{\sf ul}, j}^{\alpha/2}}, \
g_{ij} = \frac{\hat{g}_{ij}}{d_{ ij}^{\alpha/2}} \notag
\end{align}
for $i, j \in [1:J]$, $n \in [1:M_{\sf d}]$, and $m \in [1:M_{\sf u}]$.
Here, $\hat{h}_i(n)$, $\hat{f}_j(m)$, and $\hat{g}_{ij}$ are i.i.d. fading components drawn from $\mathcal{CN}(0,1)$ and
$d_{{\sf dl}, i}$, $d_{{\sf ul}, j}$, and $d_{ij}$ are the distance between the BS and the $i$th DL user, the distance between the BS and the $j$th UL user, and 
the distance between the $j$th UL user and the $i$th DL user respectively.
In the same manner, channel coefficients related to inter-cell interfering links can be defined.

For comprehensive comparison, we also consider the impacts of self-interference and user scheduling. 
As the same reason in Section \ref{subsec:sum_rates},we assume that the residual self-interference power is assumed to be the same as the noise power and regarded it as noise in simulation for the proposed schemes.
As a consequence, inter-cell user-to-user interference and BS-to-BS interference occur for the proposed schemes due to the FD operation at BSs while they do not appear in the HD systems.

For the HD systems, one DL user out of the $J$ DL users is scheduled in each cell for both no CSIT and the partial CSIT models.
For the proposed schemes, on the other hand, one DL user out of the $J$ DL users is scheduled for no CSIT model, while $K_{\sf d}$ DL users out of the $J$ DL users are scheduled for the partial CSIT model.
For both the HD systems and the proposed schemes, $K_{\sf u}$ UL users out of the $J$ UL users are scheduled. 

Fig. \ref{rate_symmetric2} plots the average sum rates of the proposed schemes and the HD systems with respect to $J$ when $K_{\sf d} = K_{\sf u} = M_{\sf d} = M_{\sf u} = 2$, $\alpha = 3$,  and $P_{\sf ref} = 10$ dB. We consider round-robin and max-SNR algorithms for user scheduling.
As seen in Fig. \ref{rate_symmetric2}, the proposed schemes outperform the conventional HD systems when both round-robin and max-SNR scheduling are used,
which attributes to the fact that inter-cell user-to-user interference is aligned into the same signal subspace where intra-cell user-to-user interference is aligned for the proposed schemes, so that inter-cell user-to-user interference is also cancelled out when removing intra-cell user-to-user interference.
Furthermore, Fig. \ref{rate_symmetric2} shows that the rate gap between the proposed schemes and the HD systems with max-SNR scheduling increases as $J$ increases,
while the rate gap with round-robin scheduling remains unchanged regardless of $J$ and is marginal compared to the case with max-SNR scheduling,
which demonstrates that in conjunction with interference management techniques, user scheduling in FD cellular networks might improve the sum rate further compared to the conventional HD systems.
From the simulation results, both user scheduling algorithms and interference management techniques for suppressing inter-cell interference are indispensable for applying FD radios into multicell cellular networks.

\section{Conclusion} \label{sec:conclusion}
In this paper, we studied the sum DoF of FD cellular networks consisting of a FD BS, HD DL users and HD UL users.
In particular, we completely characterized the sum DoF of FD cellular networks for no CSIT model 
and established an achievable sum DoF for the partial CSIT model.
Our results demonstrated that reconfigurable antennas only at the FD BS can improve the sum DoF and eventually double the sum DoF as both the numbers of DL and UL users and preset modes increase in the presence of user-to-user interference. 
We further demonstrated that such DoF improvement yields the sum rate improvement compared to the conventional HD cellular networks at the finite SNR regime.
Beyond this work, the impact of multiple reconfigurable antennas at FD BSs will be a promising future research topic.

\appendix
\section*{Proof of Lemma \ref{lemma:dl2}}
In this appendix, we prove Lemma \ref{lemma:dl2}.
%
First, we show that $\operatorname{rank}(\mathbf{P})=L_{\sf d}n_{\sf d}$ almost surely.
Recall that $\bar{\alpha}_2(t) = (t-1)|L_{\sf d} + 1$ for $t \in [1 : L_{\sf d} L_{\sf u}]$. 
Let us permute the columns of $\mathbf{P}$ as in the following order and denote the resultant matrix as $\mathbf{A}$:
\begin{align*}
\{1, 1+L_{\sf d}, \cdots, 1+(L_{\sf u}-1)L_{\sf d},  \ 2, 2+L_{\sf d}, \cdots, 2+(L_{\sf u}-1)L_{\sf d}, 
 \ \cdots, \  L_{\sf d}, 2L_{\sf d},\cdots, L_{\sf u}L_{\sf d}\}.
\end{align*}
From the definition of $\bar{\alpha}_2(t)$, $\mathbf{A} \in\mathbb{C}^{L_{\sf d} n_{\sf d} \times L_{\sf d} L_{\sf u} }$ is then given by
\begin{align}
 \mathbf{A} = \left[
\begin{array}{ccc}
h_1(1) \mathbf{W}_3^H (\mathbf{I}_{L_{\sf u}} \otimes \mathbf{e}_{L_{\sf d}}(1)) &  \cdots & h_1(L_{\sf d}) \mathbf{W}_3^H (\mathbf{I}_{L_{\sf u}} \otimes \mathbf{e}_{L_{\sf d}}(L_{\sf d}))\\
\vdots &  \ddots & \vdots \\
h_{L_{\sf d}}(1) \mathbf{W}_3^H (\mathbf{I}_{L_{\sf u}} \otimes \mathbf{e}_{L_{\sf d}}(1)) & \cdots & h_{L_{\sf d}}(L_{\sf d}) \mathbf{W}_3^H (\mathbf{I}_{L_{\sf u}} \otimes \mathbf{e}_{L_{\sf d}}(L_{\sf d}))\\
\end{array}
\right]. \notag 
\end{align}
%

Let $\mathbf{A}_i = \mathbf{W}_3^H (\mathbf{I}_{L_{\sf u}} \otimes \mathbf{e}_{L_{\sf d}}(i)) \in \mathbb{C}^{ n_{\sf d} \times L_{\sf u}}$ for $i \in [1:L_{\sf d}]$.
Since any submatrix of the IDFT matrix is a full-rank matrix \cite{bader2007petascale} and $n_{\sf d} \leq L_{\sf u}$, $\operatorname{rank}(\mathbf{A}_i)=n_{\sf d}$ so that it is right invertible.
Denoting the right inverse matrix of $\mathbf{A}_i$ by $\mathbf{A}_i^{\dagger}=\mathbf{A}_i^H ( \mathbf{A}_i \mathbf{A}_i^H)^{-1}$, 
the following relation holds:
\begin{align}
\mathbf{A} \operatorname{diag}(\mathbf{A}_1^{\dagger}, \cdots, \mathbf{A}_{L_{\sf d}}^{\dagger}) = 
\underbrace{
\left [
\begin{array}{ccc}
h_{1}(1) &  \cdots & h_{1}(L_{\sf d})\\
\vdots &  \ddots & \vdots \\
h_{L_{\sf d}}(1) & \cdots & h_{L_{\sf d}}(L_{\sf d})
\end{array}
\right ]}_{ \triangleq \mathbf{H} \in \mathbb{C}^{L_{\sf d} \times L_{\sf d}} } \otimes  \ \mathbf{I}_{n_{\sf d}} \in \mathbb{C}^{L_{\sf d} n_{\sf d} \times L_{\sf d} n_{\sf d}}. \notag
\end{align}
Since every element in $\mathbf{H}$ is i.i.d. drawn from a continuous distribution, $\mathbf{H}$ is a full-rank matrix almost surely so that $\operatorname{rank}(\mathbf{H}\otimes   \mathbf{I}_{n_{\sf d}})=L_{\sf d} n_{\sf d}$ almost surely.
Because $\operatorname{rank}(\mathbf{A})\geq \operatorname{rank}(\mathbf{H}\otimes   \mathbf{I}_{n_{\sf d}})$, we finally have $\operatorname{rank}(\mathbf{P}) = \operatorname{rank}(\mathbf{A}) \geq  L_{\sf d} n_{\sf d}$ almost surely. 
Obviously, $\operatorname{rank}(\mathbf{P}) \leq L_{\sf d} n_{\sf d}$ from the dimension of $\mathbf{P}$.
Therefore, $\operatorname{rank}(\mathbf{P}) = L_{\sf d} n_{\sf d}$ almost surely.

Next, we show that $\operatorname{rank}(\mathbf{Q}) \geq L_{\sf u} n_{\sf u}$ almost surely.
Recall that $\bar{\beta}_2(t) = (t-1)|L_{\sf u} + 1$ for $t \in [1 : L_{\sf d} L_{\sf u}]$.
Let $\mathbf{Q}_{\sf sub} =  \left [ \mathbf{F}_1(\bar{\beta_2}) \mathbf{W}_4,\cdots, \mathbf{F}_{L_{\sf u}}(\bar{\beta_2}) \mathbf{W}_4 \right ]\in \mathbb{C}^{L_{\sf d} L_{\sf u} \times L_{\sf u} n_{\sf u} }$, which is a submatrix of $\mathbf{Q}$.
In the following, we will show that $\operatorname{rank}(\mathbf{Q}_{\sf sub}) = L_{\sf u} n_{\sf u}$ almost surely, which guarantees that $\operatorname{rank}(\mathbf{Q}) \geq L_{\sf u} n_{\sf u}$ almost surely.
Let us permute the columns of $\mathbf{Q}_{\sf sub}^T $ as in the following order and denote the resultant matrix as $\mathbf{B}$:
\begin{align*} 
\{1, 1+L_{\sf u}, \cdots, 1+(L_{\sf d}-1)L_{\sf u}, \ 2, 2+L_{\sf u}, \cdots, 2+(L_{\sf d}-1)L_{\sf u}, \ \cdots, \ L_{\sf u}, 2L_{\sf u},\cdots, L_{\sf u}L_{\sf d}\}.
\end{align*}
From the definition of $\bar{\beta}_2(t)$, $\mathbf{B} \in\mathbb{C}^{L_{\sf u}n_{\sf u}\times L_{\sf d }L_{\sf u}}$ is given by
\begin{align}
\mathbf{B} =  \left[
\begin{array}{cccc}
f_1(1) \mathbf{W}_4^T (\mathbf{I}_{L_{\sf d}} \otimes \mathbf{e}_{L_{\sf u}}(1)) 
 & \cdots 
& f_1(L_{\sf u}) \mathbf{W}_4^T (\mathbf{I}_{L_{\sf u}} \otimes \mathbf{e}_{L_{\sf d}}(L_{\sf u}))\\
\vdots  & \ddots & \vdots \\
f_{L_{\sf u}}(1) \mathbf{W}_4^T (\mathbf{I}_{L_{\sf d}} \otimes \mathbf{e}_{L_{\sf u}}(1)) 
& \cdots & f_{L_{\sf u}}(L_{\sf u}) \mathbf{W}_4^T (\mathbf{I}_{L_{\sf d}} \otimes \mathbf{e}_{L_{\sf u}}(L_{\sf u}))\\
\end{array}
\right].\notag 
\end{align}

Let $\mathbf{B}_i = \mathbf{W}_4^T (\mathbf{I}_{L_{\sf d}} \otimes \mathbf{e}_{L_{\sf u}}(i)) \in \mathbb{C}^{n_{\sf u} \times L_{\sf d}}$ for $i \in [1:L_{\sf u}]$.
Since every submatrix of IDFT matrix is full-rank \cite{bader2007petascale} and $n_{\sf u} \leq L_{\sf d}$, $\mathbf{B}_i$ is a full-rank and right invertible matrix. 
Denoting the right inverse matrix of $\mathbf{B}_i$ as $\mathbf{B}_i^{\dagger}= \mathbf{B}_i^H (\mathbf{B}_i \mathbf{B}_i^H)^{-1}$, the following relation holds:
\begin{align}
\mathbf{B} \operatorname{diag}\left (\mathbf{B}_1^{\dagger}, \cdots, \mathbf{B}_{L_{\sf u}}^{\dagger} \right ) = 
\left [
\begin{array}{cccc}
f_{1}(1) &  \cdots & f_{L_{\sf u}}(1)\\
\vdots &  \ddots & \vdots \\
f_{1}(L_{\sf u}) & \cdots & f_{L_{\sf u}}(L_{\sf u})
\end{array}
\right ]^T \otimes \mathbf{I}_{n_{\sf u}} \in\mathbb{C}^{L_{sf u}n_{\sf u}\times L_{\sf u}n_{\sf u}}. \notag
\end{align}
Then, $\operatorname{rank}(\mathbf{Q}_{\sf sub}(\bar{\beta}_2)) = \operatorname{rank}(\mathbf{B}) = L_{\sf u} n_{\sf u}$ almost surely, which completes the proof of Lemma \ref{lemma:dl2}.

\bibliographystyle{IEEEbib}
\bibliography{IEEEabrv,ref}

\end{document}